\documentclass[10pt,conference]{IEEEtran}
\usepackage{cite}
\usepackage{amsmath,amssymb,amsfonts}
\usepackage{algorithmic}
\usepackage{graphicx}
\usepackage{textcomp}
\usepackage[hyphens]{url}
\usepackage{fancyhdr}
\usepackage{hyperref}
\usepackage{booktabs}
\usepackage{multirow}
\usepackage{subfigure}
\usepackage{circledtext}
\usepackage{array}
\usepackage[table]{xcolor}
\usepackage{comment}

\newcommand{\hpcayear}{2026}

\newcommand{\hpcasubmissionnumber}{NaN}

\newpage
\title{GyRot: Leveraging Hidden Synergy between Rotation and Fine-grained Group Quantization for Low-bit LLM Inference}

\clearpage
\def\hpcacameraready{} 

\newcommand\hpcaauthors{Sangjin Kim, Yuseon Choi, Byeongcheol Kim, Jungjun Oh and Hoi-jun Yoo}
\newcommand\hpcaaffiliation{School of Electrical Engineering, Korea Advanced Institue of Science and Technology (KAIST)}
\newcommand\hpcaemail{\textit{\{sangjinkim, yuseon.choi, bc27.kim, ojj1245, hjyoo\}}@kaist.ac.kr}

\author{
  \ifdefined\hpcacameraready
    \IEEEauthorblockN{\hpcaauthors{}}
      \IEEEauthorblockA{
        \hpcaaffiliation{} \\
        \hpcaemail{}
      }
  \else
    \IEEEauthorblockN{\normalsize{HPCA \hpcayear{} Submission
      \textbf{\#\hpcasubmissionnumber{}}} \\
      \IEEEauthorblockA{
        Confidential Draft \\
        Do NOT Distribute!!
      }
    }
  \fi 
}

\fancypagestyle{camerareadyfirstpage}{%
  \fancyhead{}
  
  \fancyhead[C]{
    \ifdefined\aeopen
    \parbox[][12mm][t]{13.5cm}{\hpcayear{} IEEE International Symposium on High-Performance Computer Architecture (HPCA)}    
    \else
      \ifdefined\aereviewed
      \parbox[][12mm][t]{13.5cm}{\hpcayear{} IEEE International Symposium on High-Performance Computer Architecture (HPCA)}
      \else
      \ifdefined\aereproduced
      \parbox[][12mm][t]{13.5cm}{\hpcayear{} IEEE International Symposium on High-Performance Computer Architecture (HPCA)}
      \else
      \parbox[][0mm][t]{13.5cm}{\hpcayear{} IEEE International Symposium on High-Performance Computer Architecture (HPCA)}
    \fi 
    \fi 
    \fi 
    \ifdefined\aeopen 
      \includegraphics[width=12mm,height=12mm]{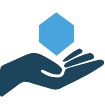}
    \fi 
    \ifdefined\aereviewed
      \includegraphics[width=12mm,height=12mm]{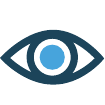}
    \fi 
    \ifdefined\aereproduced
      \includegraphics[width=12mm,height=12mm]{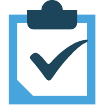}
    \fi
  }
  \fancyfoot[C]{}
}
\begin{document}
\maketitle

\ifdefined\hpcacameraready 
  \thispagestyle{camerareadyfirstpage}
  \pagestyle{empty}
\else
  \thispagestyle{plain}
  \pagestyle{plain}
\fi

\newcommand{\hpcaheight}{0mm}
\ifdefined\eaopen
\renewcommand{\hpcaheight}{12mm}
\fi


\begin{abstract}
Low-bit quantization is essential for efficient LLM inference, and both rotation and fine-grained group quantization have shown individual promise. However, their combination often leads to accuracy degradation or hardware overhead due to a mismatch between the global nature of rotation and the localized behavior of group scaling. We propose GyRot, a quantization framework and hardware accelerator that bridges this gap through algorithm–hardware co-design. GyRot introduces Coarse Rotation, Fine Grouping (CoRFiG) and Harmonic-Aligned Permutation (HAP) to enable cooperative integration of rotation and group quantization, enhancing quantizability while relaxing scaling factor precision. To further reduce hardware cost, we reformulate asymmetric quantization and introduce a zero-point rounding strategy that enables fully integer dequantization. Implemented on an INT4-based tensor PE architecture, GyRot achieves state-of-the-art 4-bit accuracy across LLaMA-family models, while delivering up to 3.4× speedup and 3.6× energy efficiency over baseline LLM accelerators. These results validate GyRot’s practical effectiveness for scalable and energy-efficient LLM deployment.
\end{abstract}

\section{Introduction}

Large Language Models (LLMs)\cite{llama, llama2, llama3} have demonstrated breakthrough performance in various natural language understanding and generation tasks. However, their massive parameter count and computational intensity impose significant inference costs, especially in edge and datacenter environments with stringent latency, energy, and memory constraints. To address these challenges, low-bit quantization\cite{gptq, rptq, quarot, spinquant, duarot, duquant, dfrot, atom, smoothquant, awq, qserve, mx, mxe, amxfp, mant} has emerged as a promising solution by compressing weights and activations to lower arithmetic precision.

Two of the most effective approaches for enhancing quantization accuracy are \textbf{rotation-based quantization}\cite{quarot, spinquant, duarot, duquant, dfrot} and \textbf{group quantization}\cite{gptq, atom, mx, mxe, amxfp, mant, awq, zeroquant_v2, nvfp4}. Rotation-based quantization employs orthogonal transformations, such as Hadamard matrices, to redistribute outliers and flatten directional variance, thereby improving quantizability. In contrast, group quantization divides channels into smaller groups, applying scaling factors and biases per group, which effectively balances numerical accuracy and hardware efficiency. Notably, recent trends have favored finer group sizes (16–64 channels)\cite{mx, mxe, amxfp, edgediff, nvfp4, mant, zeroquant_v2}, moving away from the coarser sizes (128–1024 channels) used initially \cite{gptq, atom, awq}.

Despite their individual advantages, naïvely combining rotation and group quantization often leads to non-cooperative interactions, especially at finer group sizes. Rotation inherently disperses outliers across channels, while group quantization thrives on localized scaling; merging these conflicting behaviors undermines scale coherence and increases quantization error. Recent studies, such as AMXFP~\cite{amxfp}, empirically validate this issue, showing accuracy deterioration when rotation is combined with small group sizes (e.g., 32 channels). 

Moreover, fine-grained group quantization itself introduces frequent floating-point operations during dequantization. This overhead further escalates when asymmetric quantization is adopted, as it also requires per-group zero-point handling.

In this paper, we pose a fundamental question:  
\emph{Can rotation and fine-grained group quantization be made cooperative—and if so, how can their synergy be effectively unlocked at both the algorithm and hardware levels?}

To answer this, we introduce GyRot, an algorithm-hardware co-design solution that effectively integrates rotation, group quantization, and asymmetric quantization for accurate and efficient low-bit LLM inference. We first identify that the primary cause of quantization degradation is the conflicting nature of rotation, which disperses outliers globally, and group quantization, which captures distribution locality. To resolve this misalignment, we propose precisely controlling the rotation scope, making it local enough to preserve group-level coherence. Moreover, by leveraging the harmonic characteristics of rotation matrices, we enforce locality even across groups, enabling effective integration of rotation with fine-grained group quantization under asymmetric quantization settings. This optimization also unveils hardware-level opportunities, such as reducing the required precision of scaling factors and zero-points, which significantly lowers the overhead of dequantization.

We propose GyRot with the following three contributions:\begin{itemize}
\item We propose \textbf{CoRFiG (Coarse Rotation, Fine Grouping)}, a novel method that applies rotation locally but at a coarser granularity compared to the quantization group size. This preserves local variances within quantization groups while still providing the benefits of distribution flattening. 
\item To further enhance this synergy, we introduce \textbf{HAP (Harmonic-Aligned Permutation)}, which strategically maps outlier channels to harmonic rows in the Hadamard matrix. HAP significantly improves quantization accuracy and reduces the precision requirements for scaling factors and zero-points.

\item We reformulate asymmetric quantization to effectively mitigate the precision overhead caused by long-tailed zero-point distributions. Additionally, by carefully designing a zero-point rounding strategy, we eliminate zero-point-induced clipping errors with minimal complexity, substantially alleviating the precision requirements of zero-points.

\item We implement the \textbf{GyRot accelerator}, a systolic-array-based inference engine featuring fully integer-based dequantization enabled by these algorithmic optimizations. Consequently, our design efficiently supports fine-grained group quantization (e.g., group size of 32) without incurring excessive floating-point overhead, achieving high throughput and energy efficiency while delivering state-of-the-art accuracy at 4-bit precision.
\end{itemize}

Experimental results demonstrate that GyRot outperforms state-of-the-art rotation and group quantization schemes in both accuracy and hardware efficiency. Across a range of LLM inference benchmarks, GyRot delivers higher quantization accuracy than prior algorithms such as Quarot~\cite{quarot} and SpinQuant~\cite{spinquant}, while our GyRot accelerator achieves a 1.42–3.40× speedup and 1.20–3.64× energy savings compared to recent designs like MANT~\cite{mant}, LightRot~\cite{lightrot}, and Tender~\cite{tender}.


\section{Background on LLM Quantization}
\subsection{Conventional LLM Quantization}
Quantizing LLMs is particularly challenging due to the prevalence of outliers in activation and weight distributions. While outlier-aware techniques have been explored in earlier neural networks~\cite{olaccel}, the issue has become more pronounced in transformer-based NLP models, which exhibit highly long-tailed distributions~\cite{bert, roberta}.

Initial approaches targeted fine-grained outliers at the element level, processing them separately to reduce quantization error~\cite{gobo, ant, olive}. In more recent LLMs, however, outliers tend to appear across channels~\cite{opt, llama, llama2, llama3, gemma}, and addressing these channel-wise outliers is now widely adopted as the standard approach. For instance, Atom~\cite{atom} identifies outlier channels and applies mixed-precision quantization by assigning higher bit-widths to them. In contrast, methods such as SmoothQuant~\cite{smoothquant} and AWQ~\cite{awq} multiply each channel by an offline-determined scale to suppress inter-channel variance in input activations or weights, thereby improving quantizability.

\subsection{Rotation-based Quantization}
Rotation-based quantization offers an alternative approach to distribution flattening by redistributing outliers across channels through a rotation transformation. A rotation matrix (e.g., the Hadamard matrix~\cite{quarot}) is applied to the input activation, effectively spreading the impact of large-magnitude values and reducing kurtosis, thereby making the input more quantizable. Notably, due to the rotation-invariance property of matrix multiplication, the inverse rotation can be fused into the weights, which guarantees computational equivalence, as formally demonstrated in~\cite{quarot}.

Hadamard matrices, commonly constructed using Sylvester's method~\cite{quarot}, provide a recursive and hardware-efficient way to generate orthogonal transformations. Starting from the base matrix, larger matrices can be constructed as follows:
\begin{equation}
H_1 = 
\begin{bmatrix}
1 & 1 \\
1 & -1
\end{bmatrix},
\quad
H_{n+1} = 
\begin{bmatrix}
H_n & H_n \\
H_n & -H_n
\end{bmatrix}
\end{equation}

This recursive structure allows for efficient computation using the Fast Hadamard Transform (FHT). Instead of directly multiplying a large rotation matrix $H_n$ with $O(n^2)$ operations, FHT requires only $O(n \log_2 n)$ operations~\cite{quip}. These properties enable efficient online rotation when required. Although parts of the rotation can be fused offline into the weights, online rotation remains necessary in layers with nonlinear operations (e.g., embedding, activation), making FHT a practical solution to minimize runtime cost~\cite{quip, quarot, lightrot}.

Some recent works adopt trainable rotation matrices to better fit the data distribution and improve quantization quality~\cite{spinquant}. Other methods explore multi-stage strategies that combine global, local, and permutation-based rotations for greater flexibility and quantizability~\cite{duarot, duquant}.

In contrast, LightRot~\cite{lightrot} introduces a hardware-motivated approach by applying local rotation to reduce rotation cost. To compensate for its limited flattening effect, it permutes outlier channels to align with the all-ones row in the Hadamard matrix, improving the quantization range when combined with asymmetric quantization.

\subsection{Group Quantization}
Group quantization offers an alternative approach to suppressing outliers by applying a per-group scale and bias. This structure localizes the influence of outliers within smaller regions, thereby reducing their impact on quantization error and enabling accurate inference even at low bit widths.

Formally, for a group \( g \), and bit-width \( b \), group quantization can be expressed as:
\begin{equation}
\hat{x}_i = \text{clip}\left( \left\lfloor \frac{x_i}{s_x} \right\rceil,\ q_{\min},\ q_{\max} \right), \quad s_x = \frac{2 \cdot \max(|x_g|)}{2^b - 1}
\end{equation}
\label{sec:equation}
\textcolor{black}{where $g$ indexes a quantization group, and $x_g=\{x_i \mid i\in g\}$ denotes the set of activation values within group $g$,} \( s_x \) is the scaling factor computed from the input activation values in group \( g \), and \( q_{\min} = -2^{b-1} \), \( q_{\max} = 2^{b-1} - 1 \) are the lower and upper bounds for signed \( b \)-bit quantization (e.g., \([-8, 7]\) for 4-bit).

Given group-wise quantized input \( \hat{x} \) and weight \( \hat{w} \), the inner product is reconstructed as:
\begin{equation}
y \approx \sum_{g \in \mathcal{G}} s_x^{(g)} s_w^{(g)} \cdot \sum_{i \in g} \hat{x}_i \cdot \hat{w}_i,
\end{equation}
where \( \mathcal{G} \) denotes the set of all groups, and \( s_x^{(g)} \), \( s_w^{(g)} \) are the scaling factors for input and weight in group \( g \), respectively.

Due to its relative efficiency and hardware-friendliness, group quantization has become a standard strategy in modern LLM quantization pipelines~\cite{vsquant, msfp, bfp, mx, nvfp4}. The accuracy benefit of group quantization increases as the group size becomes smaller, enabling finer-grained suppression of channel variation. Consequently, while earlier methods typically used large groups (e.g., 128 or 256) to reduce scaling overhead, recent works have demonstrated that smaller groups (e.g., 32, 16, or even 8) can yield significantly better accuracy~\cite{mx, amxfp, nvfp4, microscopiq}. Despite the increased hardware cost, this trend has been validated and adopted even in industry-grade formats~\cite{msfp, mx, nvfp4}, demonstrating its practical effectiveness.

\subsection{Asymmetric Quantization}
Asymmetric quantization is particularly effective when the data distribution is skewed, as it enables non-zero centering through the use of a zero-point. This allows for better utilization of the representational range compared to symmetric quantization. As a result, many algorithms and hardware implementations adopt asymmetric schemes to improve accuracy~\cite{smoothquant, zeroquant_v2, atom, omniquant, qserve, quarot, spinquant, resq, flatquant, ostquant}.

Like symmetric quantization, asymmetric schemes can be applied at various granularities, such as per-tensor, per-channel, or per-group. However, in addition to scaling factors, a zero-point must also be stored and applied for each unit of granularity, which introduces non-negligible overhead at finer scales. Consequently, many designs use asymmetric quantization only at the per-tensor~\cite{panacea} or per-channel~\cite{smoothquant, omniquant, qserve, resq, flatquant, ostquant} level to reduce metadata and computation cost.

Nonetheless, group-wise asymmetric quantization~\cite{amxfp, afpq, lightrot, atom} has also been explored due to its compatibility with group quantization. Because asymmetry is often more pronounced in small local groups than across entire channels or tensors, applying asymmetric quantization at the group level can significantly improve accuracy. For example, AMXFP \cite{amxfp} demonstrated that this combination yields a synergistic effect.

The group-wise asymmetric quantization of input activation can be expressed as:
\begin{align}
\hat{x}_i &= \text{clip}\left( \left\lfloor \frac{x_i}{s_x} \right\rceil + z_x,\ q_{\min},\ q_{\max} \right), \nonumber \\
s_x &= \frac{\max(x_g) - \min(x_g)}{2^b - 1}, \quad
z_x = \left\lfloor -\frac{\min(x_g)}{s_x} \right\rceil
\label{eq:asym_quant_orig}
\end{align}
where \( s_x \) and \( z_x \) are the scale and zero-point for group \( g \), and \( q_{\min} = 0 \), \( q_{\max} = 2^b - 1 \) for \( b \)-bit asymmetric quantization (e.g., \([0, 15]\) for 4-bit).

While this approach improves accuracy, the additional \( z_x \) term in Equation~(4) increases computational cost during inference. For this reason, it is rarely applied to both input and weight simultaneously \cite{amxfp}. Instead, many implementations apply asymmetric quantization to either activation~\cite{panacea, lightrot, smoothquant, atom, qserve, quarot, spinquant, resq, flatquant, ostquant} or weight~\cite{bitmod} only.

When applied only to input activations (with symmetric quantization for weights), the inner product is reconstructed as:
\begin{equation}
\begin{aligned}
y &\approx \sum_{g \in \mathcal{G}} s_x^{(g)} s_w^{(g)} \cdot \sum_{i \in g} \left( \hat{x}_i - z_x^{(g)} \right) \cdot \hat{w}_i \\
&=\sum_{g \in \mathcal{G}} s_x^{(g)} s_w^{(g)} \cdot \left(\sum_{i \in g}  \hat{x}_i\cdot \hat{w}_i - z_x^{(g)} \cdot \sum_{i \in g}  \hat{w}_i \right)
\end{aligned}
\end{equation}
This contrasts with the symmetric case in Equation~(3), where no bias term is involved.

Asymmetric quantization is also used in floating-point quantization, where schemes such as~\cite{afpq, amxfp} avoid explicit zero-point terms by assigning separate scaling factors for positive and negative values. However, as discussed in~\cite{amxfp}, when both input and weight are quantized in this way, the computing unit must handle four distinct scale combinations based on the sign of each operand, which increases computational and control complexity.

\section{Motivation}
Rotation, fine group quantization, and asymmetric quantization have each proven effective in improving the accuracy and efficiency of LLM inference under low-bit constraints. In this section, we analyze the interplay between these quantization techniques from both model accuracy and hardware efficiency perspectives, and highlight several key insights that motivate the design of our proposed method.

\subsection{Model Accuracy Perspective}

\textbf{Observation 1: Asymmetric quantization can be synergistic with fine group quantization and rotation.}  
Recent studies have shown that smaller group sizes in group quantization lead to increasingly skewed value distributions. For instance, AMXFP~\cite{amxfp} quantitatively demonstrates that per-group activation distributions become more asymmetric as the group size decreases, highlighting the benefit of applying asymmetric quantization at finer granularities. Furthermore, LightRot~\cite{lightrot} shows that optimized rotation—while improving quantizability—exacerbates distributional asymmetry by redistributing outliers across channels. This skewness can be effectively mitigated through asymmetric quantization. These findings indicate that asymmetric quantization not only complements fine-grained group quantization, but also works synergistically with rotation-based transformations.

\begin{figure}[]
\includegraphics[width=3.6in]{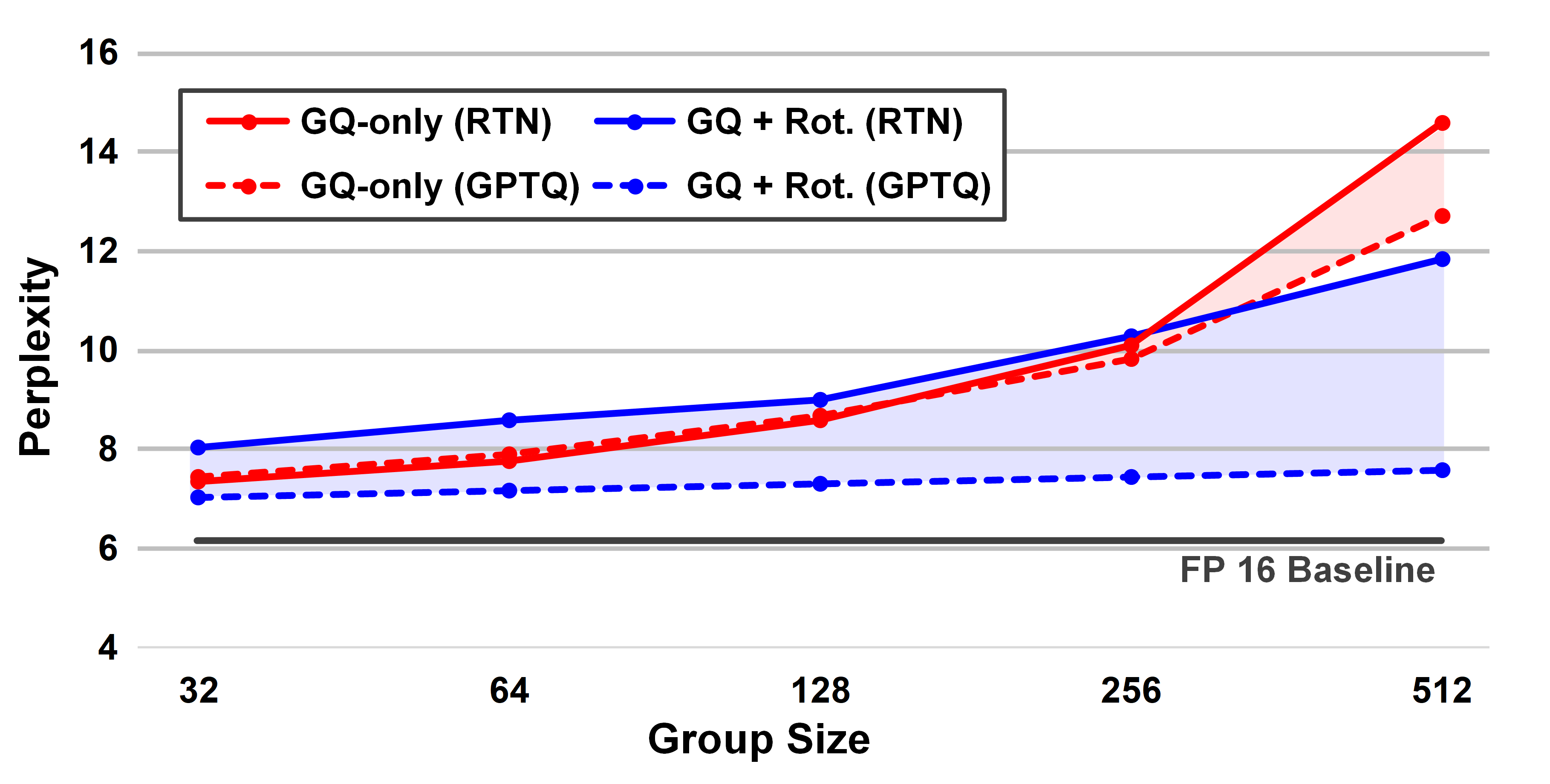}
\centering
\caption{Effect of data rotation with different quantization granularities. Perplexity measured with LLaMa3-8B \cite{llama3} on WikiText-2 dataset \cite{wikitext2}}
\vspace{-6pt}
\label{fig:f1}
\end{figure}
\textbf{Observation 2: Fine-grained group quantization is non-cooperative with rotation.}  
Fig.~\ref{fig:f1} illustrates the quantization accuracy as a function of group size using both round-to-nearest (RTN) and GPTQ~\cite{gptq} quantizers. For large group sizes, applying rotation clearly provides accuracy benefits. However, as the group size decreases, group quantization alone significantly improves perplexity, while combining it with rotation yields little to no improvement. This divergence becomes even more pronounced under the RTN baseline, where the error compensation effect of GPTQ is absent: at smaller group sizes, perplexity is actually reversed, leading to worse accuracy when rotation is applied.
Similar findings are reported in AMXFP~\cite{amxfp}, where applying rotation with a group size of 32 resulted in accuracy degradation, ultimately leading to the removal of rotation. This discrepancy stems from a fundamental mismatch between the two techniques: rotation globally redistributes values across all channels to flatten the overall distribution, while group quantization is designed to capture local variations across groups. This implies that there is still room for improvement when combining the two techniques, if their distinct optimization characteristics are carefully taken into account.

\subsection{Hardware Cost of Group Quantization}



\begin{figure}[]
\includegraphics[width=3.4in]{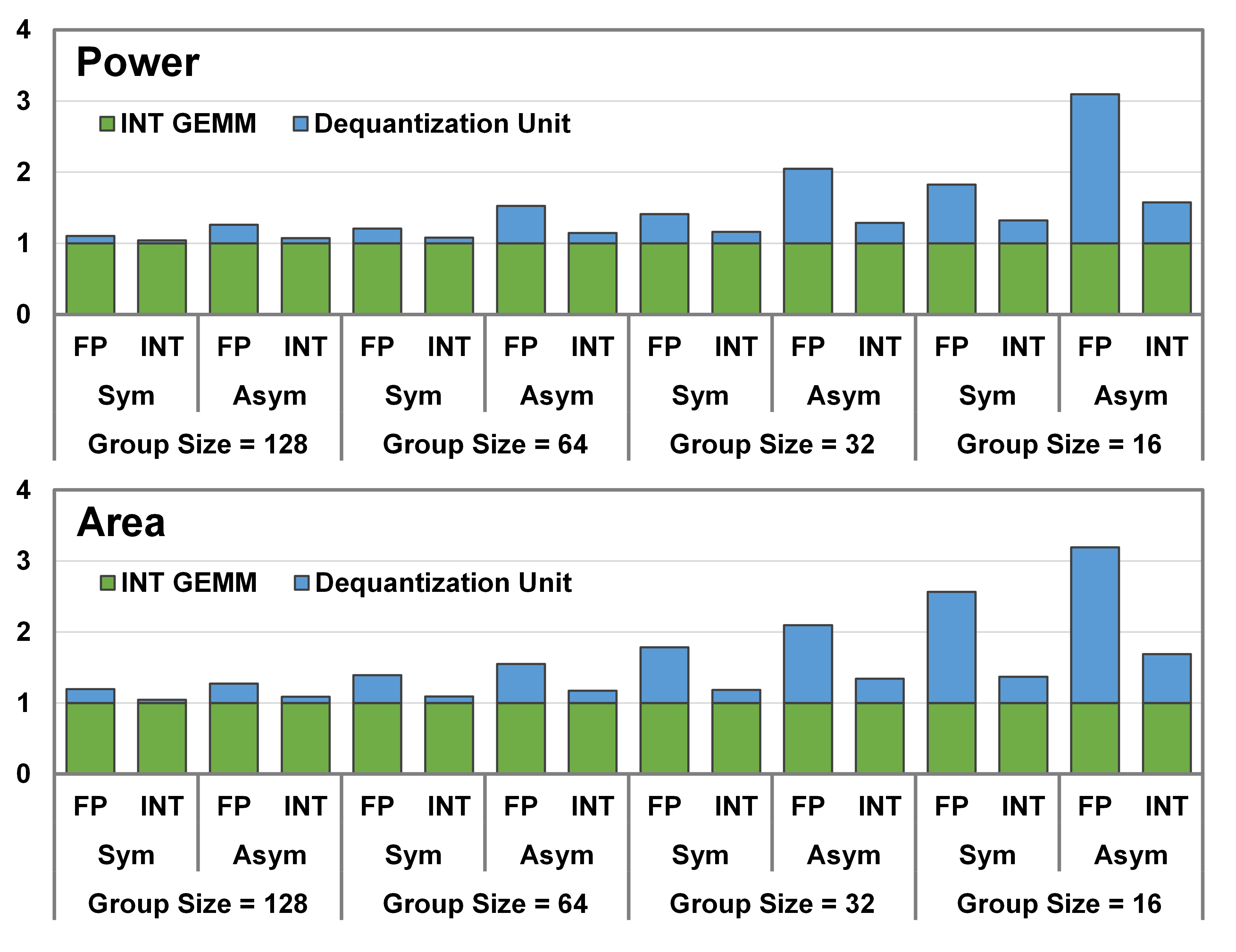}
\centering
\caption{\textcolor{black}{Hardware cost with different quantization granularity.}}
\vspace{-6pt}
\label{fig:f2}
\end{figure}

\textbf{Observation 3: Smaller group sizes increase hardware cost, further amplified by asymmetric quantization.} 
\label{sec:observation3}
Using smaller group sizes is effective in reducing quantization error, but it also leads to a larger number of groups, resulting in more frequent dequantization operations. While intra-group MAC operations are typically performed using low-bit integer arithmetic (e.g., INT4), \textcolor{black}{\emph{the dominant source of overhead is the floating-point dequantization datapath} used by prior designs to preserve accuracy when applying per-group scales and zero-points.} 
Fig.~\ref{fig:f2} separates INT GEMM from the dequantization unit and reports both \emph{FP} and \emph{INT} dequantization paths. As the group size decreases, the cost of the \emph{FP} dequantization datapath grows sharply; this overhead becomes even larger with asymmetric quantization due to the additional zero-point term. \textcolor{black}{By contrast, the \emph{INT} dequantization path (with INT8 scale/zero-point) remains much smaller, motivating our fully-integer design in Sec.~V.}

Based on these three observations, there is a clear need for a method that can effectively combine rotation, group quantization, and asymmetric quantization—each individually beneficial—for more efficient and accurate low-bit LLM inference.
To address this challenge, we propose \textbf{GyRot}, an algorithm–hardware co-design solution that leverages the hidden synergy between rotation and fine-grained group quantization under an asymmetric quantization framework. In particular, GyRot explores cooperative optimization strategies between rotation and fine-grained group quantization to achieve higher accuracy, while also relaxing the precision requirements of scaling factors and zero-points. \textcolor{black}{As a result, GyRot quantizes both the scale factor and zero-point to INT8 and applies them inside the PE as a fully integer dequantization path.} This design choice results in reduced dequantization overhead and improved hardware efficiency.

\section{Algorithm-level Optimization}

\begin{figure*}[]
\includegraphics[width=6.8in]{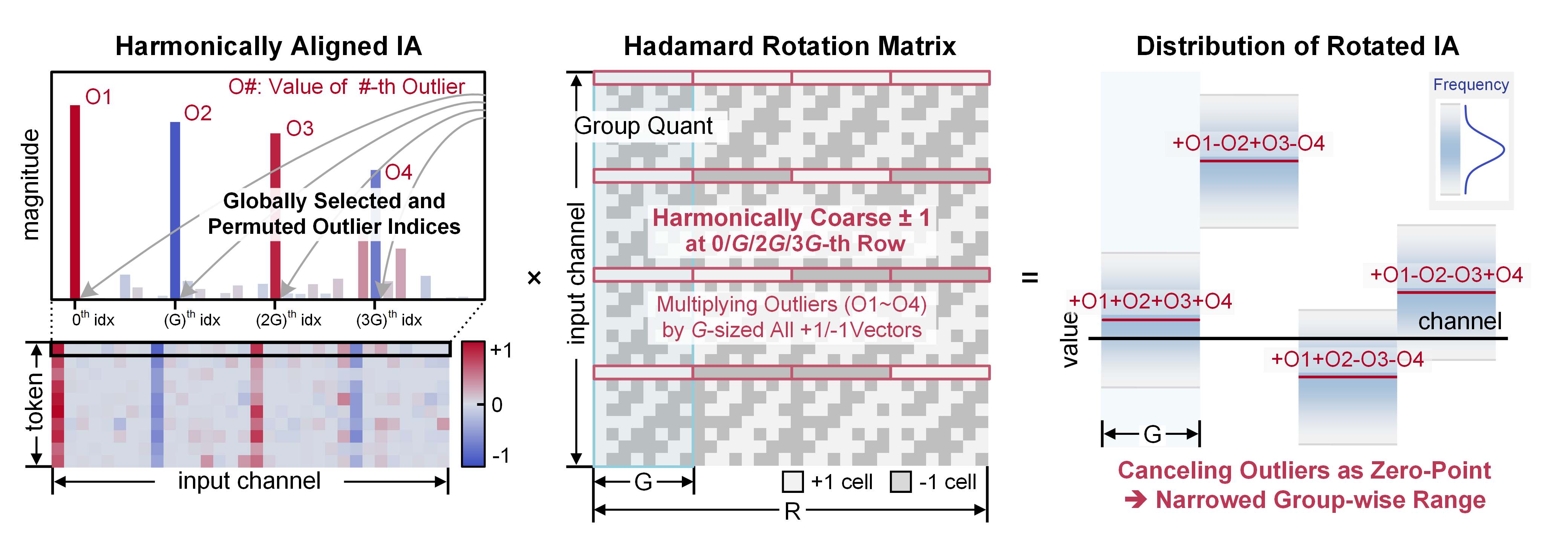}
\centering
\caption{\textcolor{black}{Proposed quantization algorithm: Coarse-Rotation, Fine Grouping (CoRFiG) with Harmonic-Aligned Permutation (HAP). (G = 8, R = 32 case.)}} 
\label{fig:f3}
\vspace{-6pt}
\end{figure*}

\begin{figure}[]
\includegraphics[width=\linewidth]{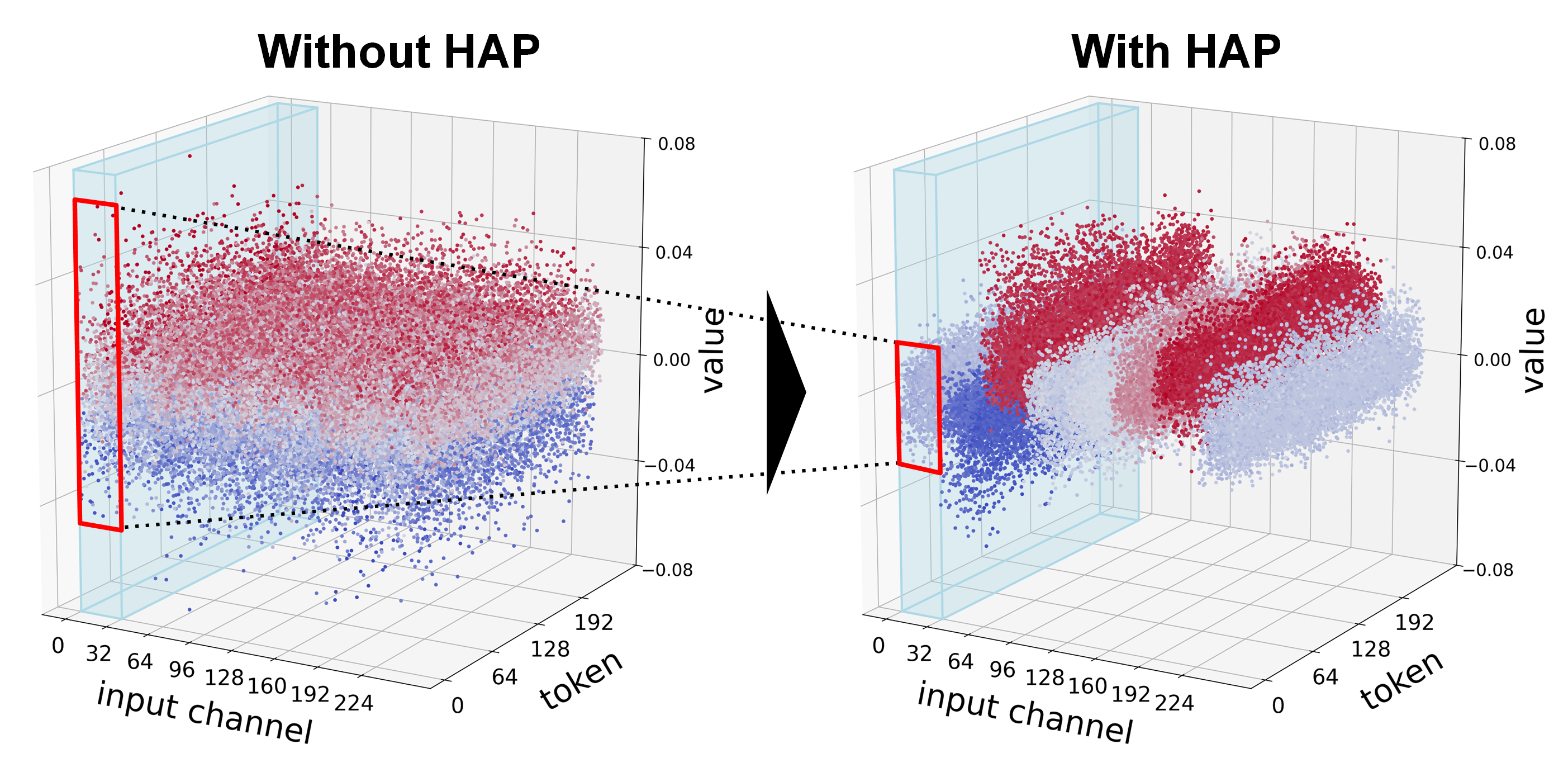}
\centering
\caption{Effect of HAP on activation distribution after rotation.} 
\label{fig:f4}
\vspace{-6pt}
\end{figure}

\subsection{Rotation for Fine-grained Group Quantization}
Rotation and fine-grained group quantization are non-cooperative by default because they take opposing approaches to improving quantizability.
Rotation applies a transformation (e.g., a rotation matrix) to activations or weights, globally redistributing values across all channels. This operation naturally \textbf{\textit{amortizes the influence of outliers across the entire tensor}}, thereby flattening the distribution and enhancing quantizability.
In contrast, group quantization preserves the original structure of the data but partitions it into smaller groups, each quantized independently using its own scaling factor (and possibly zero-point). This approach \textbf{\textit{isolates the impact of outliers within each group}}, allowing for finer-grained adaptation to local distribution variations.
Therefore, while group quantization focuses on containing outlier effects locally, rotation aims to spread them globally—revealing a fundamental mismatch that \textit{undermines their compatibility} when applied together.

Building on this insight, we propose \textbf{Coarse-Rotation Fine-Grouping (CoRFiG)} and \textbf{Harmonic-Aligned Permutation (HAP)}, as illustrated in Fig.~\ref{fig:f3}, which tailor the rotation strategy to better cooperate with fine-grained group quantization.

Instead of applying global rotation across the entire channel dimension, CoRFiG performs rotation locally within a specified rotation scope \( R \), where \( R = 2^r < N_{ch} \) for a positive integer \( r \). To preserve the flattening benefits of rotation, CoRFiG chooses a sufficiently large \( R \)—referred to as coarse rotation. Specifically, it maintains the relation \( R = 2^g\cdot G \), where \( G \) is the group size used for quantization and \( g \) is a positive integer.
This design enables effective redistribution of outliers within the local scope \( R \), achieving a flattened distribution and enhanced quantizability, while limiting the spread of each outlier to only \( R \) channels rather than the entire channel dimension. At the same time, by keeping the group size \( G \) small, CoRFiG preserves the benefits of fine-grained group quantization. This coarse–fine decoupling balances the trade-off between outlier amortization and localized adaptation.
Our evaluation in Section~VI primarily focuses on the configuration \( R = 1024 \), \( G = 32 \).

While CoRFiG ensures sufficient flattening within \( R \) channels and isolates the influence of outliers across rotation scopes, HAP further improves quantizability by aligning group-wise ranges using the harmonic characteristics of the Hadamard matrix.
As described in Equation~(1), Hadamard matrices are recursively constructed. A Hadamard matrix of size \( 2^{n+1} \times 2^{n+1} \), denoted \( H_{n+1} \), is composed of two \( H_n \) matrices: the top half as \([H_n, H_n]\) and the bottom half as \([H_n, -H_n]\). Due to this recursive structure, Hadamard matrices contain repeating “harmonically coarse” \( \pm1 \) vectors—i.e., length-\( 2^k \) vectors of all \( +1 \) or all \( -1 \)—at regular strides of \( 2^k \) for \( k < n \), forming structured harmonic patterns.

As illustrated in Fig.~\ref{fig:f3}, HAP leverages these harmonic rows to separate the range of each group after rotation. For example, with \( G = 8 \) and \( R = 32 \), we have \( R = 2^g \cdot G \) with \( 2^g = 4 \), implying that there are four coarse harmonic rows (G, 2G, 3G, 4G-th row). By permuting globally selected high-magnitude outlier channels (\( O_1 \sim O_4 \)) to align with these harmonic rows prior to rotation, each outlier is multiplied with a consistent sign (all \( +1 \) or all \( -1 \)) within its group.  As a result, unlike the unaligned case where outliers are randomly mixed with both \( +1 \) and \( -1 \), HAP produces group-wise distributions that are tightly bounded with shifted biases, as shown in Fig.~\ref{fig:f4}. This group-wise range reduction not only reduces quantization error but also significantly lowers the precision requirements of the scaling factors used for each group, thereby reducing dequantization overhead.

However, while distribution within each group becomes narrower, the post-rotation distribution within each group becomes considerably asymmetric, potentially increasing the precision requirements of zero-points. In this case, the central value of each group is determined by linear combinations of the rotated outlier channels (e.g., \( +O_1 + O_2 + O_3 + O_4 \), \( +O_1 - O_2 + O_3 - O_4 \), etc.). To mitigate this, CoRFiG uses a sufficiently large rotation scope \( R \) relative to group size \( G \), and additional asymmetric quantization optimizations are applied.

\subsection{Further Optimizing Asymmetric Quantization}
While the CoRFiG+HAP combination effectively improves quantizability and mitigates scaling factor precision requirements, the resulting group-wise distributions often become highly biased and asymmetric. However, under such highly asymmetric distributions, conventional asymmetric quantization suffers from an expanded zero-point range, which limits hardware efficiency. In this section, we reformulate asymmetric quantization to better align with the characteristics of CoRFiG-HAP and adjust the rounding policy to further enhance robustness.

{\bf Reformulating Asymmetric Quantization.} 
\begin{figure}[]
\includegraphics[width=3.4in]{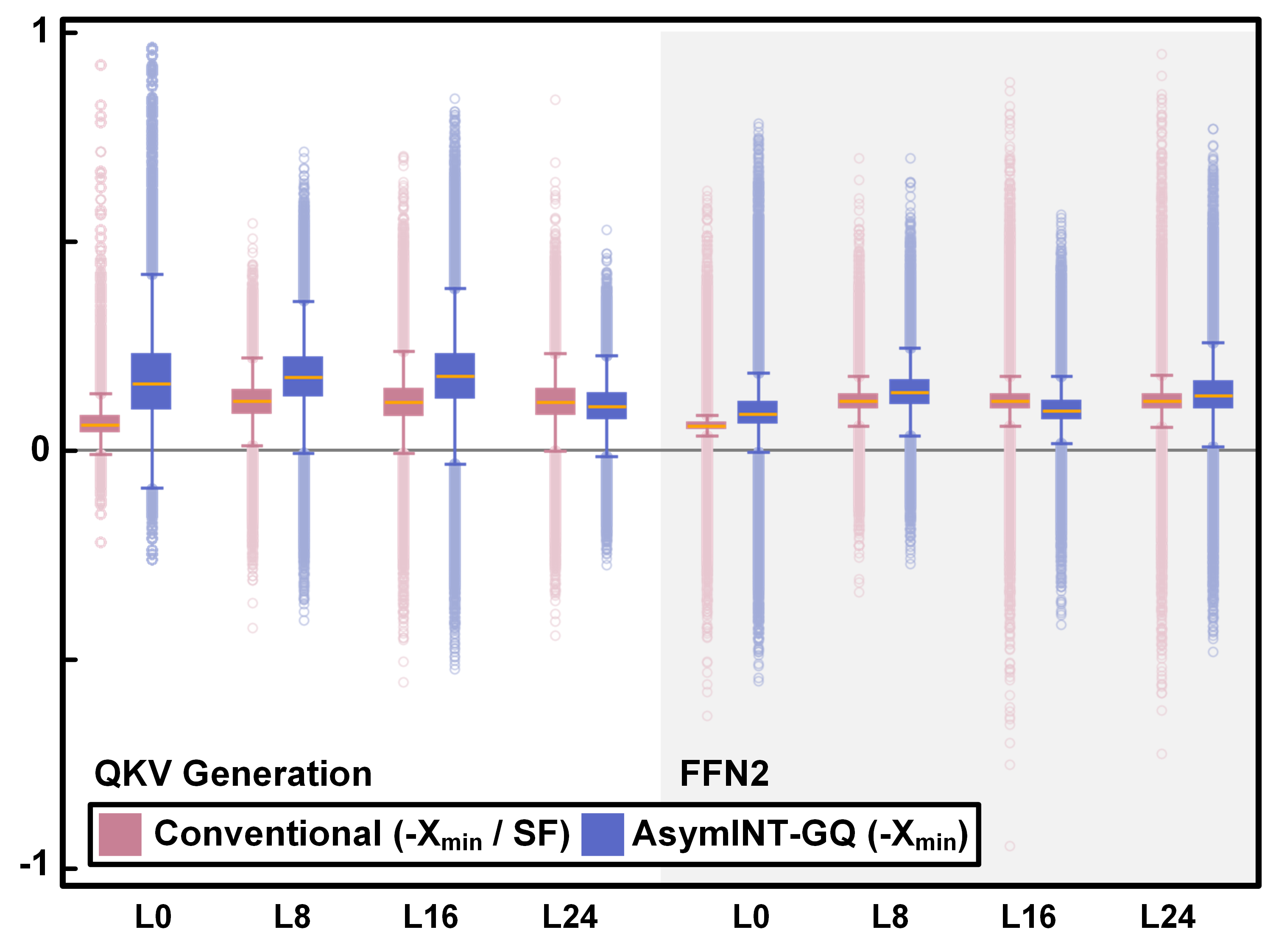}
\centering
\caption{Comparison of zero-point distributions in conventional and reformulated quantization. The boxes represent the second and third quartiles and the median, the whiskers indicate the 1\% and 99\% percentiles, and the circle represents outliers. zero-points are normalized with per-layer power-of-two scale.}
\label{fig:f5}
\vspace{-6pt}
\end{figure}
As explained in Equation~(4), conventional asymmetric quantization performs scaling first, followed by zero-point biasing: \( \hat{x} = \left\lfloor x / s_x + z_x \right\rceil \). In this formulation, the zero-point is defined in the scaled domain as \( z_x = -\min(x_g)/s_x \). For typical activation distributions—where the degree of asymmetry is relatively mild compared to the full dynamic range—the resulting zero-point values tend to stay within a narrow range. However, in cases of highly asymmetric activations—particularly under HAP—this formulation can produce extremely long-tailed zero-point distributions due to the small values of \( s_x \).

To address this issue, we reformulate the quantization procedure by reversing the order of zero-point biasing and scaling. Specifically, we define the quantized value as \( \hat{x} = \left\lfloor (x + z_x) / s_x \right\rceil \), where the zero-point is computed directly from the unscaled domain as \( z_x = -\min(x_g) \). This reformulation avoids division during zero-point calculation, thereby mitigating the amplification effect caused by small scaling factors.

As illustrated in Fig.~\ref{fig:f5}, the proposed formulation yields significantly flatter zero-point distributions. The figure presents box plots of normalized zero-points across eight LLaMA-3-8B layers, including QKV and FFN components. Unlike the conventional method, which exhibits long-tailed distributions with narrow box ranges and prominent outliers, our method produces wider box ranges, indicating shorter tails and lower kurtosis.

\begin{figure}[]
\includegraphics[width=2.8in]{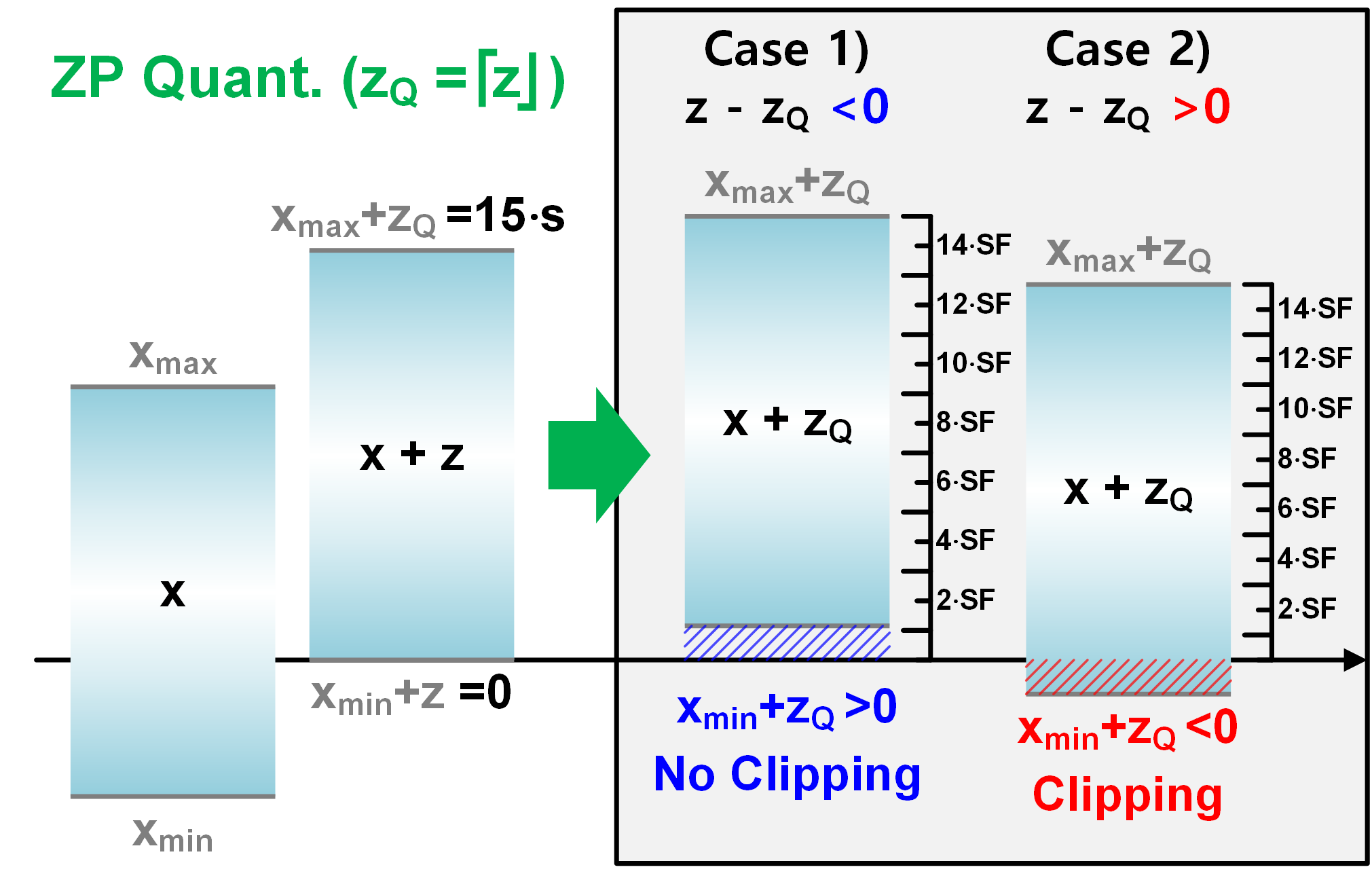}
\centering
\caption{Effect of zero-point quantization according to the sign of quantization error.}
\label{fig:f6}
\vspace{-12pt}
\end{figure}
\textbf{Rethinking the Rounding Strategy for Zero-Point.}  
In addition to reformulation, we analyze the impact of the zero-point rounding strategy on quantization error. Asymmetric quantization shifts the data range by adding a zero-point \( z_x \), mapping the minimum value to 0 and the maximum to \( 2^b - 1 \). However, quantizing \( z_x \) itself introduces a rounding error \( \delta_z = z - z_Q \), which affects the placement of the minimum value, as illustrated in Fig.~\ref{fig:f6}.  

If \( \delta_z \le 0 \), the shifted minimum \( x_{\min} + z_Q \ge 0 \), and no clipping occurs—although some portion of the quantization range is wasted. In contrast, if \( \delta_z > 0 \), then \( x_{\min} + z_Q < 0 \), resulting in range underflow and data clipping, which leads to significant quantization error.

To avoid this clipping behavior, we replace the conventional \textit{rounding} operation with a \textit{ceiling} function when quantizing the zero-point. This guarantees \( z_Q \ge z \), ensuring that the shifted minimum remains within the valid range and eliminating the risk of underflow at the lower bound.

\textbf{Final Formula.} The reformulated group-wise asymmetric quantization is summarized as:
\begin{align}
\hat{x}_i &= \text{clip}\left( \left\lfloor \frac{x_i + z_x}{s_x} \right\rceil,\ q_{\min},\ q_{\max} \right), \nonumber \\
z_x &= \left\lceil -\min(x_g) \right\rceil,\quad
s_x = \frac{\max(x_g) + z_x}{2^b - 1}.
\label{eq:asym_quant}
\end{align}

This formulation also modifies the dequantization process used during matrix multiplication. When both weights and activations are quantized, the inner product is computed as:
\begin{equation}
y \approx \sum_{g \in \mathcal{G}} s_w^{(g)} \cdot \left(s_x^{(g)} \cdot \sum_{i \in g}  \hat{x}_i\cdot \hat{w}_i - z_x^{(g)} \cdot \sum_{i \in g}  \hat{w}_i \right)
\end{equation}

Compared to the conventional formulation in Equations~(4) and (5), the only change lies in the order of operations: the scaling factor and zero-point are applied in reverse order. Note that we apply a \textit{ceiling} function when computing the zero-point to ensure range safety and prevent underflow during quantization.

\begin{figure*}[]
\centering
\subfigure[]{
\includegraphics[width=1.9in]{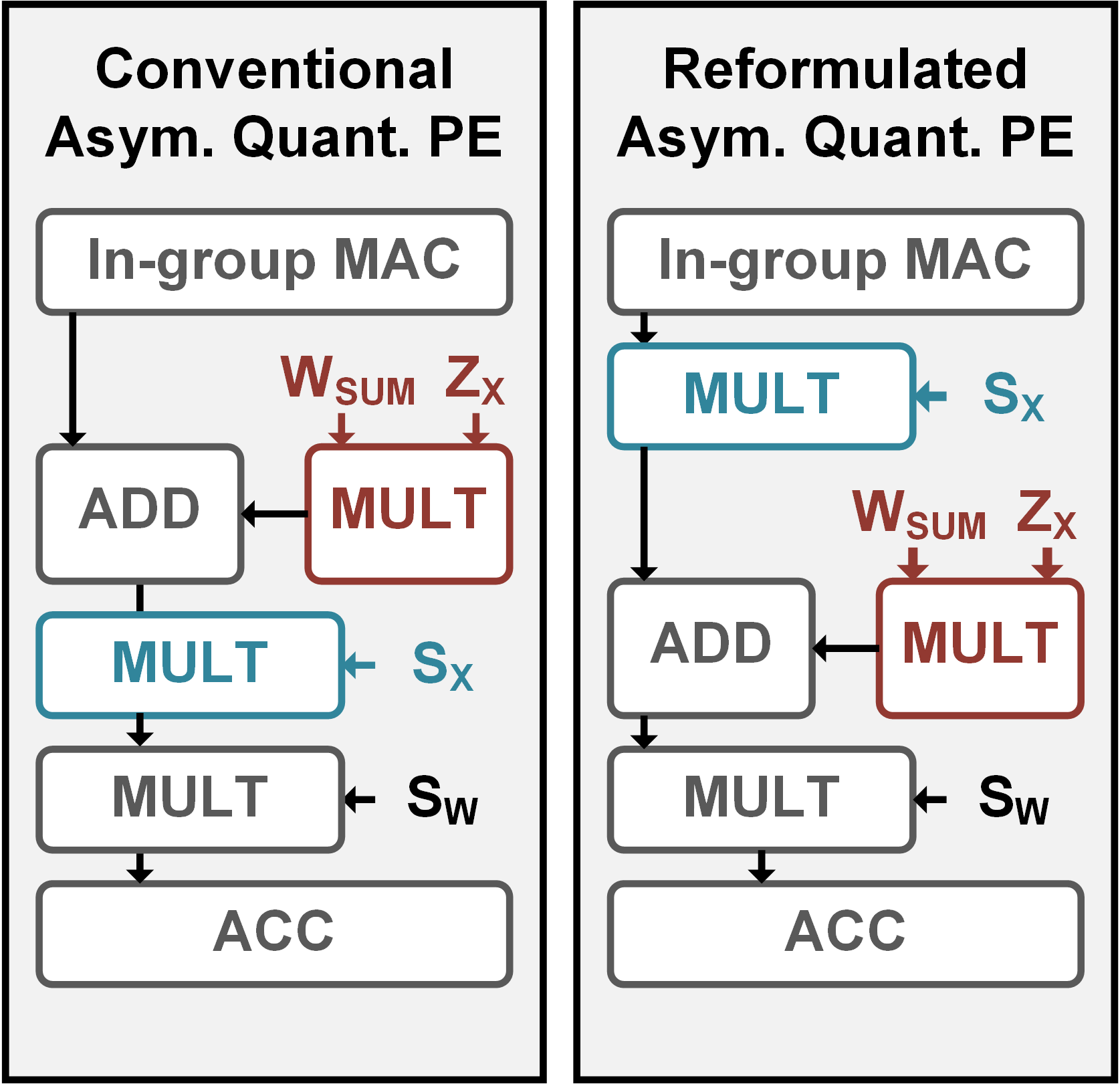}
\centering
}
\subfigure[]{
\includegraphics[width=3.8in]{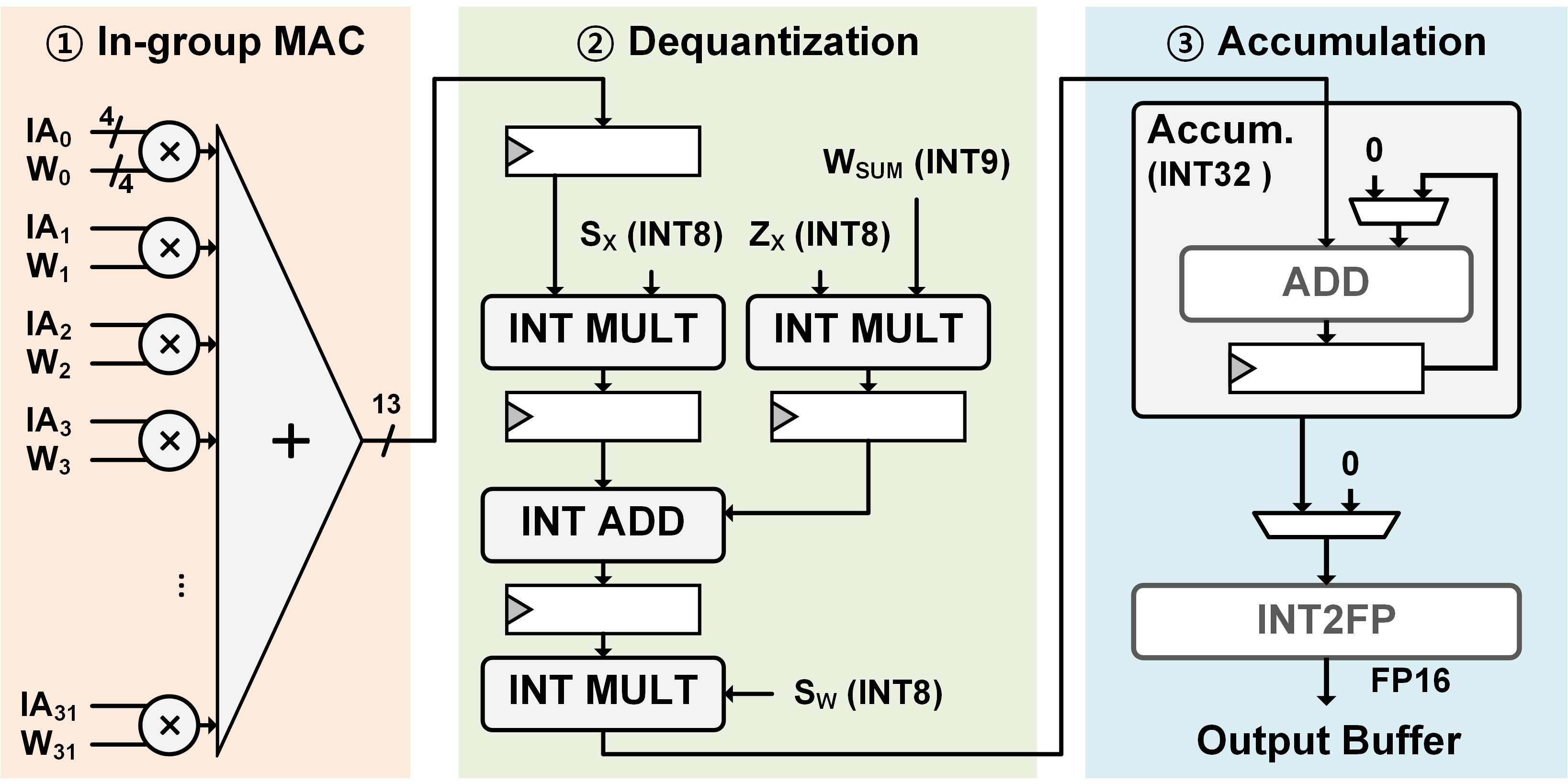}
\centering
}
\caption{GyRot PE. (a) Operation flow change for reformulated asymmetric quantization (Equa. \ref{eq:asym_quant}).(b) Microarchitecture.}
\label{fig:f8}
\vspace{-6pt}
\end{figure*}

In summary, we reformulate asymmetric quantization to align with CoRFiG-HAP, which introduces severe group-level asymmetry through localized outlier alignment.

\section{GyRot Microarchitecture}
This section outlines the microarchitectural details of {\it GyRot} that enable efficient and accurate low-bit LLM inference. Our accelerator integrates architectural components that support the combined use of rotation and group quantization, as described in Sec. IV-B. It also supports efficient computation with asymmetrically quantized input, as detailed in Sec. IV-C.

\subsection{Processing Element for Reformulated Asymmetric Quant.}
The use of asymmetric and group quantization requires large dequantization operations and inter-group accumulation. Moreover, as the group size decreases, the hardware cost (in area and energy) of dequantization becomes significant. By leveraging CoRFiG, HAP, and reformulated asymmetric quantization, {\it GyRot} utilizes a fully integer dequantization datapath with integer-quantized scaling factors and zero-points.
\label{sec:pe_structure}
As shown in Fig.~\ref{fig:f8}(a), the reformulated dequantization operation derived from Equation~\ref{eq:asym_quant} simply changes the order of operations: the activation scaling factor $S_X$ is applied first, followed by the addition of the zero-point term ($Z_X \times W_{\mathrm{SUM}}$). The detailed microarchitecture of the {\it GyRot} PE is depicted in Fig.~\ref{fig:f8}(b):
\textcircled{\scriptsize 1} In each cycle, the PE performs a 32-way dot product between 4-bit integer input activations ($X_{0\sim31}$) and 4-bit integer weights ($W_{0\sim31}$). This configuration supports a minimum group size of 32 under INT4 quantization, and the resulting dot product yields a 13-bit partial sum.
\textcircled{\scriptsize 2} The partial sum is then processed through the dequantization stage. The activation scaling factor $S_X$ is applied to the partial sum; concurrently, the zero-point $Z_X$ is multiplied by the precomputed group-wise weight sum $W_{\mathrm{SUM}} = \sum_{i \in g} \hat{w}_i$ and added to the result. Finally, the weight scaling factor $S_W$ is applied to complete the dequantization. All arithmetic operations are pipelined, and the associated metadata ($S_X$, $Z_X$, $S_W$) are represented with 8-bit integer precision.
\textcircled{\scriptsize 3} Since the entire computation remains within the integer domain, the resulting value can be accumulated using a 32-bit integer accumulator. Before writing the final output to the buffer, the accumulated value is converted to FP16 to reduce the output bitwidth.

\subsection{GyRot Accelerator}

\begin{figure}[]
\centering
\includegraphics[width=3.5in]{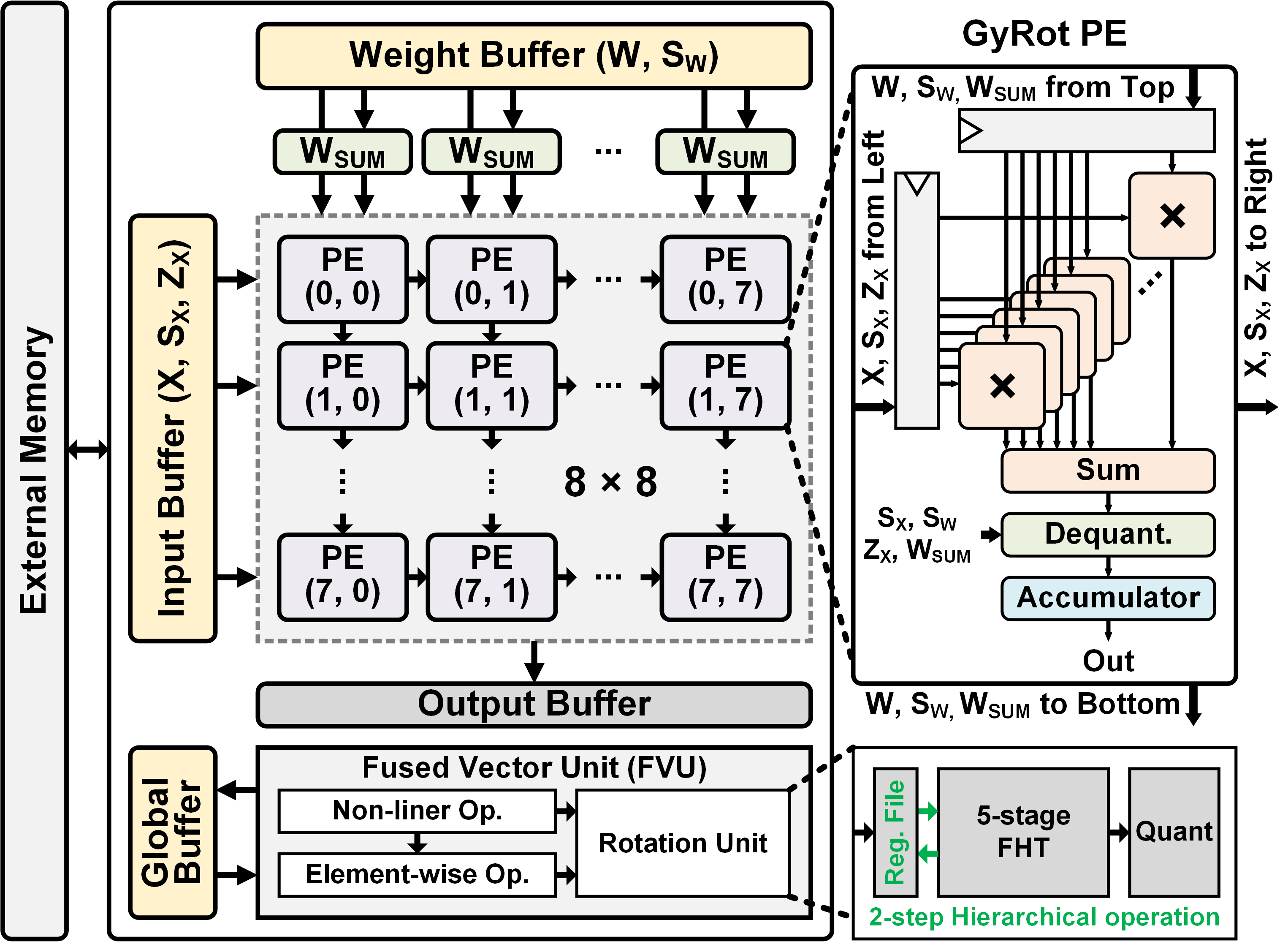}
\caption{{\it GyRot} accelerator architecture.}
\label{fig:f7}
\vspace{-12pt}
\end{figure}

Fig.~\ref{fig:f7} shows the architectural overview of the {\it GyRot} accelerator. The accelerator adopts an $8\times8$ systolic PE array, where each PE supports 32-way 4-bit dot products, with group quantization applied. This configuration allows the $8\times8\times32$ tensor array to perform 2048 operations in parallel.
The systolic array operates in an output-stationary manner. Each PE performs dot products for intra-group accumulation and applies dequantization to enable sequential inter-group accumulation.
The input buffer stores input activations and their associated quantization metadata ($X$, $S_X$, $Z_X$), while the weight buffer stores weights and their scaling factors ($W$, $S_W$). The group-wise weight sum $W_{\mathrm{SUM}}$, required for asymmetric dequantization, is computed once and shared across the entire row. A multi-bank memory structure is employed for both input and weight buffers to provide sufficient bandwidth; dedicated banks are reserved for metadata.

The Fused Vector Unit (FVU) is integrated to support the rotation operations with CoRFiG and HAP, particularly when non-linear or element-wise functions are applied in between layers.
While usual rotation operations can be fused into the weights of preceding or succeeding layers due to the rotation-invariance of matrix multiplication, certain rotations must instead be performed on-line when non-linear layers—such as self-gated activations~\cite{glu} or embedding layers~\cite{rope}—intervene between linear layers, as discussed in~\cite{quarot, spinquant}. To support such cases, a dedicated rotation and quantization unit is integrated within the FVU, enabling on-line rotation immediately after non-linear operations.
When the FVU loads output activations from global memory, it performs the nonlinear or element-wise operation before applying the rotation; the results are then directly passed to the rotation units for fused execution. The rotation unit implements the Hadamard rotation using a fast Hadamard transform (FHT)~\cite{quarot, lightrot}, requiring only $O(n\log_2 n)$ additions and subtractions for rotating a vector of length $n$.
\label{sec:fht_unit}
We implement a 5-stage, 32-way FHT unit composed of 160 add/subtract units (32 units per stage). By incorporating a local register file and executing a two-stage scheme, the unit supports scalable rotations up to $32\times 32=1024$ dimensions. \textcolor{black}{Partial gating of the FHT lanes also enables sub-32 power-of-two sizes ($2,4,8,16,32$) without wasting energy. Thus, the FHT unit supports $R$ to powers of two up to 1024, and the quantization group size $G$; under CoRFiG we choose $R=2^{g}\!\cdot\!G$ with $R\le 1024$.
}
We implement a 5-stage, 32-way FHT unit composed of 160 add/subtract units (32 units per stage). By incorporating a local register file and executing a two-stage scheme, the unit supports scalable rotations up to $32\times 32=1024$ dimensions. \textcolor{black}{Partial gating of the FHT lanes also enables sub-32 power-of-two sizes ($2,4,8,16,32$) without wasting energy. Thus, the FHT unit supports rotation scopes $R$ that are powers of two up to 1024, while the quantization group size $G$ is chosen independently; under CoRFiG, we choose $R=2^{g}\!\cdot\!G$ with $R\le 1024$.}

\label{sec:permutation}
Unlike rotation, the permutation required by HAP can be fused into the weights since both non-linear and element-wise operations are permutation-invariant. Therefore, by pre-permuting the output channels of the weight matrix, the resulting activations become naturally permuted and can be directly passed to the FVU. \textcolor{black}{Consequently, HAP introduces no additional run-time overhead, since its layer-specific permutations are pre-fused into the weights and require no online computation.}

\subsection{Need of Dedicated Hardware for GyRot Algorithm}
\label{sec:need_hw}
\textcolor{black}{
While modern GPUs provide efficient tensor cores supporting low-bit integer or floating-point matrix multiplication, group-quantized inference still requires additional per-group scaling and zero-point biasing after the GEMM operation. 
These dequantization steps are executed on CUDA cores rather than tensor cores~\cite{tensorrt_llm, qserve}, and thus are performed in floating-point precision. 
As a result, a software implementation on GPUs cannot fully exploit the efficiency of integer arithmetic, since intermediate results from integer GEMM must be converted to floating-point format for dequantization and accumulation. 
This mixed-precision execution path increases both latency and energy consumption, limiting the benefits of low-bit quantization.
}

\textcolor{black}{
In contrast, the proposed \textit{GyRot} accelerator integrates a fully-integer dequantization datapath within each processing element. 
By performing all operations—including scaling and zero-point biasing—directly in the integer domain, GyRot eliminates frequent type conversions and floating-point overhead, achieving substantial improvements in hardware efficiency. 
Therefore, dedicated hardware support is essential to realize the full advantage of the GyRot algorithm.
}

\section{Evaluation}
\subsection {Experiment Setup}
{\bf Model and datasets.}
We evaluate {\it GyRot} on three families of LLMs, including LLaMA~\cite{llama}, LLaMA-2~\cite{llama2}, and LLaMA-3~\cite{llama3}, covering a range of model architectures and sizes. Depending on the evaluation objective, we select different LLaMA families to ensure fair and appropriate comparisons. 
To assess quantization quality, we first measure the perplexity (PPL) on the WikiText-2 dataset~\cite{wikitext2}, a standard benchmark for evaluating a model's basic language generation capability. While PPL does not fully capture task-specific or conversational performance, it provides a quick and consistent metric for comparison with prior work in language modeling. For this evaluation, we include both LLaMA-1 and LLaMA-2 to maintain alignment with widely used baselines in the quantization literature.

For task-level evaluation, we conduct zero-shot inference on various benchmarks, including PIQA~\cite{piqa}, ARC-e, ARC-c~\cite{arc}, BoolQ~\cite{boolq}, HellaSwag~\cite{hellaswag}, and WinoGrande~\cite{winogrande}, which collectively test commonsense reasoning, factual understanding, and logical inference. The evaluations are performed using the LM-Evaluation-Harness framework~\cite{lm-harness}. Since these tasks require stronger generalization and reasoning capabilities, we focus on more capable models such as LLaMA-2 and LLaMA-3.
Finally, to evaluate the overall response quality and practical usefulness of quantized models, we adopt the MT-bench~\cite{mt-bench} framework, which utilizes LLM-as-a-Judge, a method that leverages strong reference models to assess human-likeness and interaction quality. For this setting, we use LLaMA-3-8B-Instruct, an instruction-tuned variant specifically optimized for dialogue-based use cases.

{\bf Quantization Method.}
We compare the quantization accuracy of \textit{GyRot} against two hardware baselines that efficiently handle dynamic scales in LLMs by adjusting quantization granularity:
\begin{itemize}
\item \textbf{Tender}: LLM accelerator that mitigates outliers by chunking activation channels and grouping them such that adjacent scaling factors differ by a 1-bit shift, which is absorbed during accumulation.
\item \textbf{MANT}: Applies group quantization using a flexible data format that supports diverse distributions. It adopts a group size of 64.
\end{itemize}

Since these baselines adopt bit-flexible architectures, we evaluate multiple design points: W4A4 for all cases, W8A8 for Tender, and W4A8 for MANT, to include high-precision configurations.

As rotation-based quantization is not yet widely adopted in hardware, we additionally compare against three algorithmic baselines and one hardware accelerator baseline, all evaluated under the W4A4 configuration to emphasize high-performance inference:

\begin{itemize}
\item \textbf{Quarot} (Algorithmic baseline): Applying Hadamard matrix as rotation matrix. It performs global rotation across all channels.
\item \textbf{SpinQuant} (Algorithmic baseline): Extends Hadamard rotation matrices into a trainable space and uses Cayley optimization to improve rotation quality.
\textcolor{black}{
\label{sec:duquant}
\item \textbf{DuQuant} (Algorithmic baseline): Uses a two-stage local rotation with an intermediate permutation that globally redistributes outlier channels to further flatten the distribution.}
\item \textbf{LightRot}: Combines rotation and asymmetric group quantization with outlier-aware permutation at a granularity of 128. (Similar to CoRFiG with $R=G=128$) While effective, it relies on floating-point zero-points, resulting in high dequantization overhead and limited scalability to smaller group sizes such as 64 or 32.
\end{itemize}

\textcolor{black}{
{\bf Relationship to LightRot.}
\label{sec:compare_lightRot}
LightRot introduces outlier direction aligning, effectively aligning prominent channels to the all-ones row of a Hadamard block under the constraint $R{=}G$. In contrast, GyRot decouples the rotation scope from the grouping granularity (CoRFiG, $R{=}2^g\!\cdot\!G$) and aligns outliers to multiple harmonic rows using HAP, which tightens per-group ranges and reduces the precision requirement of scale/zero-point. Together with our reformulated asymmetric quantization and ceiling-based ZP rounding, this enables fully-integer dequantization (INT8 SF/ZP). 
}

To ensure a fair comparison, all rotation-based baselines use dynamic, asymmetric, per-token quantization for activation values. The KV cache is quantized using asymmetric quantization with a group size of 128, and weights are symmetrically quantized using GPTQ~\cite{gptq} after applying rotation.
\textit{GyRot} is configured with a group size of 32 for both activation and weight, and 128 for the KV cache. It employs rotation with asymmetric group quantization, as detailed in Section~IV.

\newcolumntype{C}[1]{>{\centering\arraybackslash}p{#1}}

\begin{table*}[t]
\centering
\caption{\textcolor{black}{Comparison of Perplexity (PPL) and Configuration with Previous Methods.}}
\scriptsize
\label{tab:perplexity_comparison}
\begin{tabular}{l|C{1.2cm}C{1.2cm}C{1.2cm}|cc|C{0.7cm}|C{1cm}C{1cm}|C{1cm}C{1cm}}
\specialrule{1pt}{2pt}{2pt}
\multirow{2}{*}{\textbf{Method}} & \multicolumn{3}{c|}{\textbf{Precision}}  & \multicolumn{2}{c|}{\textbf{Group Quant.}} & \multirow{2}{*}{\textbf{Rot.}} & \multicolumn{2}{c|}{\textbf{LLaMA-1}} & \multicolumn{2}{c}{\textbf{LLaMA-2}} \\
& W & A & KV & SF & ZP & &\textbf{1-7B} & \textbf{1-13B} & \textbf{2-7B} & \textbf{2-13B} \\
\specialrule{1pt}{2pt}{2pt}

FP16            & 16 & 16 & 16 & -- & -- &  --           & 5.68  & 5.09  & 5.47  & 4.88 \\
\specialrule{1pt}{2pt}{2pt}
Tender          & 8-Tensor & 8-Tokens & 16  & -- & -- & --    & 5.87  & 5.28  & 5.77  & 5.09 \\
MANT            & 4-G64 & 8-G64 & 16 & FP16 & -- & --    & 5.79  & 5.20  & \textbf{5.57}  & \textbf{4.96} \\
\rowcolor[HTML]{E0F2F5} 
\textbf{GyRot-INT}  & 4-G32 & 8-G32 & 4-G128  & INT8 & INT8 & --    & \textbf{5.76}  & \textbf{5.16}  & 5.60  & 4.98 \\
\specialrule{1pt}{2pt}{2pt}
Tender          & 4-Tensor & 4-Tokens & 16 & -- & -- & --            & 23.85 & 13.68 & 36.47 & 55.08 \\
Quarot          & 4-Channel & 4-Token & 4-G128 & -- & -- & \checkmark    & 6.37 & 5.59  & 6.25  & 5.49 \\
Spinquant       & 4-Channel & 4-Token & 4-G128 & -- & -- & \checkmark    & 6.12  & 5.39  & 5.96  & 5.74 \\
\textcolor{black}{DuQuant} & 4-Channel & 4-Token & 4-G128 & -- & -- & \checkmark    & 6.18  & 5.47  & 6.08  & 5.33 \\
MANT            & 4-G64 & 4-G64 & 16  & FP16 & -- & --            & 6.09  & 5.38  & 5.92  & 5.24 \\
LightRot        & 4-G128    & 4-G128  & 4-G128 & FP16 & FP16 & \checkmark    & 5.95  & 5.27  & 5.73  & 5.08 \\
\rowcolor[HTML]{E0F2F5} 
\textbf{GyRot-FP}  & 4-G32 & 4-G32 & 4-G128  & FP16 & FP16 & \checkmark    & \textbf{5.86}  & \textbf{5.22}  & \textbf{5.67}  & \textbf{5.03} \\
\rowcolor[HTML]{E0F2F5} 
\textbf{GyRot-INT}  & 4-G32 & 4-G32 & 4-G128  & INT8 & INT8 & \checkmark    & 5.89  & \textbf{5.22}  & 5.88  & \textbf{5.03} \\

\specialrule{1pt}{2pt}{0pt}
\vspace{-12pt}
\end{tabular}
\end{table*}

\begin{table*}[t]
\centering
\caption{Zero-shot Task Accuracy Comparison of Rotation-based Quantization Methods under W4A4KV4 Configuration.}
\scriptsize
\label{tab:zero_shot_comparison}
\begin{tabular}{C{2.5cm}p{2cm}cC{1cm}C{1cm}C{1cm}C{1cm}C{1cm}C{1cm}C{1cm}C{1.5cm}}
\specialrule{1pt}{2pt}{2pt}
\multirow{2}{*}{\textbf{Model}} & \multirow{2}{*}{\textbf{Method}} & \textbf{W-A-KV} & \multicolumn{6}{c}{\textbf{Benchmark}} \\
& & \textbf{Precision} & \textbf{PIQA} & \textbf{ARC-e} & \textbf{ARC-c} & \textbf{BoolQ} & \textbf{HellaS.} & \textbf{WinoG.}  & \textbf{Avg.}\\
\specialrule{1pt}{2pt}{2pt}

\multirow{5}{*}{LLaMA-2-7B} & Full Precision    & 16-16-16   & 79.16  & 74.33  & 46.42  & 77.71 & 75.94  & 69.53 & 70.52  \\
\cmidrule{2-10}
 & Quarot            & 4-4-4      & 76.50  & 69.32  & 41.30  & 72.66 & 72.09 & 63.54  & 65.90  \\
 & Spinquant         & 4-4-4      & 76.88  & 71.08  & 40.44  & 74.40 & 73.51  & 65.82 & 67.02  \\
 & LightRot          & 4-4-4      & \textbf{78.18}  & 72.73  & 43.69  & 75.60 & 74.35  & 67.72 & 68.71  \\ 
 \rowcolor[HTML]{E0F2F5} 
 \cellcolor[HTML]{FFFFFF} & \textbf{GyRot-INT}             & 4-4-4      & 77.69  & \textbf{74.28}  & \textbf{44.54}  & \textbf{77.19} & \textbf{74.65}  & \textbf{68.98} & \textbf{69.55}  \\ 

\specialrule{1pt}{2pt}{2pt}

\multirow{5}{*}{LLaMA-2-13B} & Full Precision    & 16-16-16   & 80.63  & 77.53  & 49.15  & 80.58 & 79.39  & 71.90 & 73.20  \\
\cmidrule{2-10}
 & Quarot            & 4-4-4      & 78.84  & 73.32  & 44.37  & 77.58 & 75.73  & 68.90 & 69.79  \\
 & Spinquant         & 4-4-4      & 78.29  & 74.49  & 46.67  & 76.76 & 75.22  & 67.72 & 69.86  \\
 & LightRot          & 4-4-4      & \textbf{79.33}  & \textbf{76.47} & \textbf{48.72}  & 77.68 & 77.84  & 70.96 & 71.83       \\  
\rowcolor[HTML]{E0F2F5} 
\cellcolor[HTML]{FFFFFF} & \textbf{GyRot-INT}             & 4-4-4      & \textbf{79.33}  & 76.22  & 48.21  & \textbf{80.52} & \textbf{78.15}  & \textbf{71.82} & \textbf{72.38}       \\  

\specialrule{1pt}{2pt}{2pt}

\multirow{5}{*}{LLaMA-3-8B} & Full Precision    & 16-16-16   & 80.63  & 77.74  & 53.50  & 81.10 & 79.18  & 73.01 & 74.19  \\
\cmidrule{2-10}
 & Quarot            & 4-4-4      & 76.06  & 70.58  & 43.17  & 72.66 & 72.53  & 66.77 & 66.96  \\
 & Spinquant         & 4-4-4      & 79.16  & 73.57  & 46.33  & 76.15 & 75.43  & 68.75 & 69.90  \\
 & LightRot          & 4-4-4      & \textbf{80.25}  & \textbf{78.49}  & 49.40  & 79.66 & 76.67  & 70.09 & 72.43  \\ 
\rowcolor[HTML]{E0F2F5} 
\cellcolor[HTML]{FFFFFF}  & \textbf{GyRot-INT}             & 4-4-4      & 79.65  & 78.16  & \textbf{51.45}  & \textbf{80.67} & \textbf{77.40}  & \textbf{70.56} & \textbf{72.98}  \\ 

\specialrule{1pt}{2pt}{2pt}
\end{tabular}
\vspace{-16pt}
\end{table*}

{\bf Hardware implementation.}
We evaluate the performance and energy consumption of \textit{GyRot} compared to baseline accelerators. The processing element (PE) and all associated components are implemented in RTL using SystemVerilog, and their functionality is verified through RTL simulation. We synthesize \textit{GyRot} using Samsung's 28nm technology node with Synopsys Design Compiler~\cite{synopsys_dc}, targeting an operating frequency of 1\,GHz to match prior work~\cite{tender}. On-chip SRAMs are generated using a commercial memory compiler with the same technology node. All accelerators are evaluated under iso-compute-area constraints, taking into account both the main computation and the dequantization logic. DRAM power is estimated using the DDR4 model from the Micron DRAM Power Calculator~\cite{micron-dram}.

\subsection {Accuracy Evaluation}
{\bf Perplexity Comparison with Prior Work.}
To provide a quantitative comparison with prior implementations, Table~\ref{tab:perplexity_comparison} reports the PPL achieved under various quantization configurations. \textbf{GyRot-FP} and \textbf{GyRot-INT} represent two design points of \textit{GyRot}, scaling factor (SF) and zero-point (ZP) are represented in FP16 or INT8, respectively.

Tender achieves comparable PPL to FP16 when using W8A8, but suffers a significant drop in accuracy at W4A4. MANT with W4A8 outperforms Tender-W8A8, yet \textbf{GyRot-INT-W4A8} surpasses both, even with fully integer SF and ZP.
At the W4A4 configuration, Quarot, SpinQuant and DuQuant recover much of the accuracy loss through the use of rotation, while MANT performs even better by applying group quantization. LightRot further improves accuracy by combining asymmetric group quantization with rotation. However, \textbf{GyRot-FP} achieves the best PPL with a smaller group size, and notably, \textbf{GyRot-INT} maintains competitive accuracy even under fully integer quantization of SF and ZP.

{\bf Zero-Shot Task Evaluation on Rotation Algorithms.}
To compare rotation-based quantization schemes under low-precision settings, Table~\ref{tab:zero_shot_comparison} presents zero-shot task accuracy with a consistent 4-bit configuration (W4A4KV4).

Across all model sizes, \textit{GyRot-INT} consistently outperforms prior methods, including Quarot, SpinQuant, and LightRot, despite using fully integer SF and ZP. For example, on LLaMA-3-8B, while Quarot shows a 7.3\% accuracy drop from full-precision, \textit{GyRot-INT} narrows this gap to just 1.2\%, achieving 72.98\%. 
These results confirm that \textit{GyRot}'s quantization not only preserves perplexity but also delivers strong task-level accuracy under fully integer quantization.

\begin{figure}[t]
\centering
\includegraphics[width=3.5in]{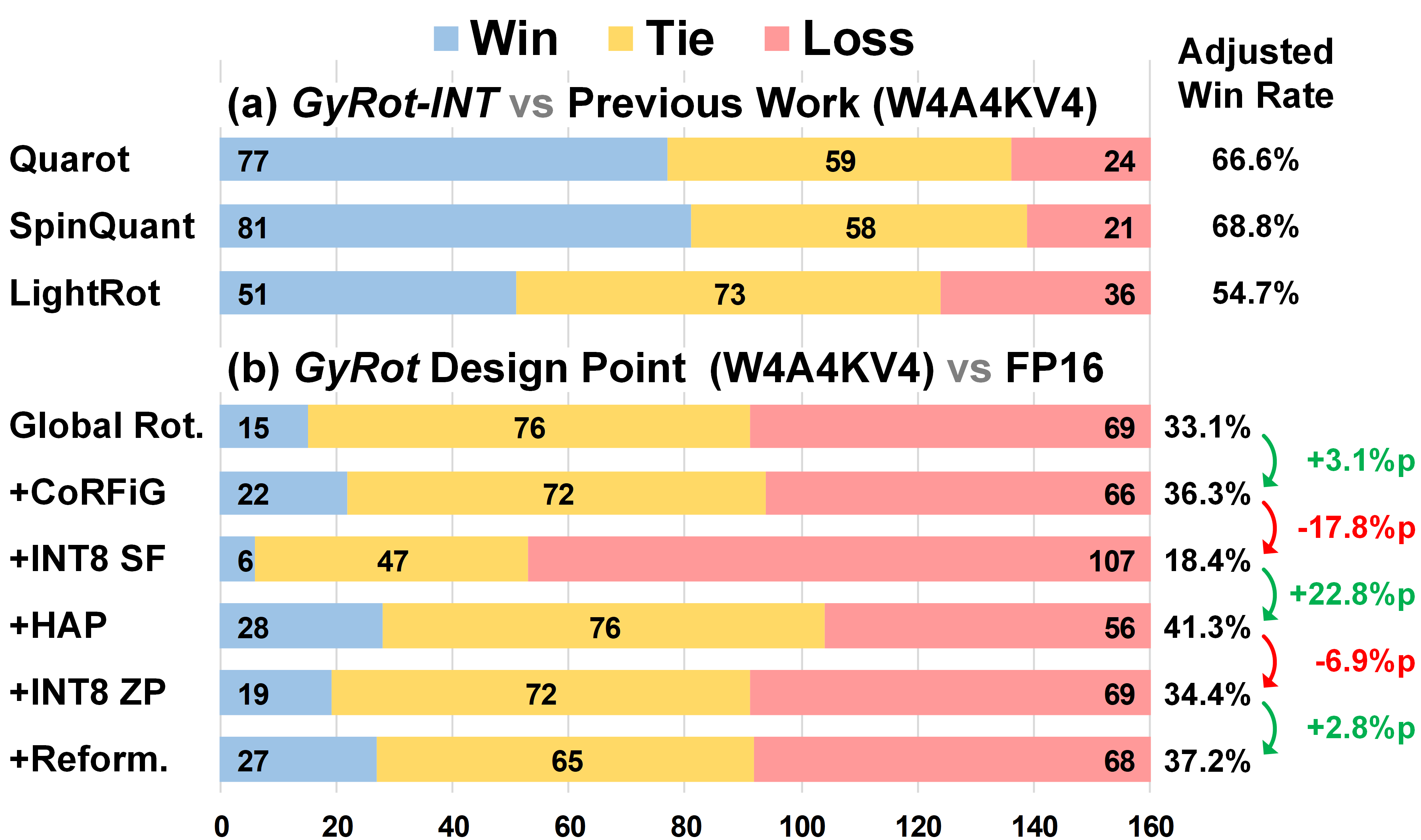}
\caption{Evaluation with LLM-as-a-Judge on MT-Bench \cite{mt-bench}, (a) Comparison with previous work (b) Progressive inclusion of our key contributions.}
\label{fig:f10}
\vspace{-12pt}
\end{figure}

{\bf Conversational Performance Evaluation with MT-Bench.}
Fig.~\ref{fig:f10}(a) compares \textit{GyRot-INT} with prior rotation-based methods under the W4A4KV4 configuration, using the MT-Bench framework with LLM-as-a-Judge across 160 turns. The adjusted win rate is computed by treating each tie as a 0.5 win. Based on this metric, \textit{GyRot} consistently outperforms previous methods, achieving 66.6\%, 68.8\%, and 54.7\% against Quarot, SpinQuant, and LightRot, respectively.

Fig.~\ref{fig:f10}(b) presents a breakdown of how each design choice in \textit{GyRot} contributes to the final performance, relative to full-precision (FP16) responses. All design points use group quantization with a group size of 32. Starting with global rotation alone, the win rate is 33.1\%, but incorporating CoRFiG improves performance by +3.1\%p by better aligning rotation scope with group quantization. In contrast, directly quantizing SF to INT8 results in a substantial drop (–17.8\%p). Adding HAP significantly recovers performance (+22.8\%p), demonstrating its ability to preserve distribution locality even after rotation. Although quantizing ZP to INT8 reduces win rate (–6.9\%p), the reformulated asymmetric quantization and ceiling-based ZP rounding recover performance, reaching 37.2\% win rate.
These results validate that each component of \textit{GyRot}—particularly CoRFiG, HAP, and the reformulated asymmetric quantization—plays a critical role in preserving generation quality under low-precision settings (W4A4) with integer-quantized SF and ZP.

{\bf Design Point Analysis of CoRFiG+HAP.}
To intuitively analyze the effects of group size and rotation size, we conduct a detailed design space exploration of \textit{GyRot-INT} on LLaMA-3-8B, as shown in Table~\ref{tab:size_analysis}. Even with global rotation, smaller group sizes yield better perplexity, highlighting the benefits of fine-grained group quantization. However, when both the group and rotation scopes are small, the ability of rotation to distribute outliers diminishes, and increased inter-group variance leads to PPL degradation under quantized SF and ZP. As the rotation size increases, PPL improves and saturates around R1024, ultimately outperforming global rotation. Based on this analysis, we select the G32–R1024 configuration as the default design point in our evaluation. These findings confirm that decoupling the granularity of rotation and group quantization enables a more flexible and accurate quantization design.

\begin{table}[]
\centering
\caption{Perplexity according to Different Group and Rotation Sizes.}
\scriptsize
\label{tab:size_analysis}
\begin{tabular}{>{\centering\arraybackslash}C{1cm}|cccccccc}
\specialrule{1pt}{2pt}{2pt}
\textbf{Model} &  \multicolumn{7}{c}{\textbf{LLaMA-3-8B}} \\
\specialrule{1pt}{2pt}{2pt}
\textbf{Group} &  \multicolumn{7}{c}{\textbf{Rotation Size}} \\
\textbf{Size} &\textbf{R32} & \textbf{R64} & \textbf{R128} & \textbf{R256} & \textbf{R512} & \textbf{R1024} & \textbf{Global}\\
\specialrule{1pt}{2pt}{2pt}

\textbf{G32}   & 30.12 & 27.83 & 7.41 & 7.36  & \textbf{6.89} & \textbf{6.91} & 7.04 \\
\textbf{G64}   & --    & 36.33 & 7.55 & 7.52  & \textbf{7.00} & \textbf{7.00} & 7.19 \\
\textbf{G128}  & --    & --    & 7.69 & 7.60  & \textbf{7.10} & \textbf{7.10} & 7.31  \\

\specialrule{1pt}{2pt}{0pt}
\end{tabular}
\vspace{-12pt}
\end{table}

{\bf Scaling Factor and Zero-Point Quantization}
To understand how CoRFiG and HAP influence the precision requirements of SF and ZP in asymmetric quantization, we analyze the perplexity sensitivity under different bit-widths. As shown in Table~\ref{tab:sf_analysis}, standard group quantization (GQ-only) suffers substantial degradation when SFs are quantized to FP8 or INT8, requiring FP16 to maintain acceptable perplexity. CoRFiG alleviates this sensitivity, narrowing the gap between FP16 and FP8. With HAP, quantization becomes notably robust—INT8 SF yields nearly identical perplexity (6.80) to FP16, demonstrating the enhanced stability of the GyRot.

Table~\ref{tab:zp_analysis} examines the impact of asymmetric quantization schemes and ZP rounding strategies on different ZP precisions. While applying CoRFiG + HAP improves perplexity, it also increases sensitivity to ZP precision. Naive rounding in the conventional asymmetric quantization leads to significant degradation under INT8 precision. Reformulating the asymmetric quantization formula narrows the gap between FP16 and INT8, and further replacing rounding with a ceiling-based strategy restores INT8 performance to near parity with FP16 (6.91 vs. 6.81). These results demonstrate that \textit{GyRot} can achieve low perplexity even with low-precision SF and ZP by carefully co-designing the quantization formulation and rounding mechanism.

\begin{table}[]
\centering
\caption{Effect of CoRFiG and HAP on scaling factor precision requirements.}
\label{tab:sf_analysis}
\scriptsize
\begin{tabular}{>{\centering\arraybackslash}m{1.8cm}|cc|ccc}
\specialrule{1pt}{2pt}{2pt}
\textbf{Model} &  \multicolumn{5}{c}{\textbf{LLaMA-3-8B}} \\
\specialrule{1pt}{2pt}{2pt}
\multirow{2}{*}{\textbf{Method}} &  
\multicolumn{2}{c|}{\textbf{Size}} & \multicolumn{3}{c}{\textbf{SF Precision}}\\
&\textbf{Group} & \textbf{Rotation} & \textbf{FP16} & \textbf{FP8} & \textbf{INT8} \\
\specialrule{1pt}{2pt}{2pt}

\textbf{GQ-only}    & 32    & --    & 7.40 & 13.38  & 416.15    \\
\specialrule{0.5pt}{2pt}{2pt}
\textbf{CoRFiG}       & 32    & 1024  & 6.91 & 7.03   & 364.17    \\
\textbf{CoRFiG+HAP} & 32    & 1024  & \textbf{6.80} & \textbf{6.84}   & \textbf{6.80}      \\

\specialrule{1pt}{2pt}{0pt}
\end{tabular}
\vspace{-12pt}
\end{table}

\begin{table}[t]
\centering
\caption{Impact of asymmetric quantization schemes and ZP rounding strategies on zero-point precision.}
\scriptsize
\label{tab:zp_analysis}
\begin{tabular}{c|cc|ccc}
\specialrule{1pt}{2pt}{2pt}
\textbf{Model} &  \multicolumn{5}{c}{\textbf{LLaMA-3-8B}} \\
\specialrule{1pt}{2pt}{2pt}
\multirow{2}{*}{\textbf{Method}} &  
\multicolumn{2}{c|}{\textbf{Asymmetric Quant.}} & \multicolumn{3}{c}{\textbf{ZP Precision}}\\
&\textbf{Formula} & \textbf{ZP Quant.} & \textbf{FP16} & \textbf{FP8} & \textbf{INT8} \\
\specialrule{1pt}{2pt}{2pt}

\textbf{GQ-only}    & --    & --    & 7.21 & 19.37  & 7.21    \\
\specialrule{0.5pt}{2pt}{2pt}
\textbf{CoRFiG} & Conv.   & Rounding  & 6.96 & 6.97   & 6.96      \\
\specialrule{0.5pt}{2pt}{2pt}
\textbf{CoRFiG+HAP} & Conv.   & Rounding  & 6.81 & 6.91   & 7.93      \\
\textbf{CoRFiG+HAP} & Reform. & Rounding  & 6.80 & 6.83   & 7.65      \\
\textbf{CoRFiG+HAP} & Reform. & Ceiling   & 6.81 & 6.83   & 6.91      \\

\specialrule{1pt}{2pt}{0pt}
\end{tabular}
\vspace{-12pt}
\end{table}

\textcolor{black}{
\label{sec:group_size_analysis}
{\bf Comparison of Rotation Algorithms under Group Quantization.}
Table~\ref{tab:group_analysis} highlights how rotation strategies interact with group quantization. Quarot (global rotation) benefits as groups get finer, but only modestly (7.31$\rightarrow$7.04). DuQuant’s two-stage rotation with global redistribution improves the per-channel case over Quarot yet shows \emph{minimal} gains as group size shrinks (8.06$\rightarrow$7.98), consistent with our analysis in Sec.~IV-A that aggressively dispersing outliers is non-synergistic with fine-grained grouping. In contrast, LightRot and GyRot—both preserving locality—achieve substantially lower PPL under FP16 SF and ZP, with LightRot slightly leading and GyRot closely matching. Crucially, when SF/ZP are quantized to INT8, LightRot’s reliance on high-precision zero-points leads to large degradation (7.69/36.33/30.12 at G=128/64/32), whereas GyRot maintains robustness (7.10/7.00/6.91) thanks to decoupled rotation scope (CoRFiG), harmonic alignment (HAP), and the reformulated asymmetric quantization.
}

\begin{table}[]
\vspace{-12pt}
\centering
\caption{\textcolor{black}{Perplexity according to Different Group Sizes.}}
\scriptsize
\label{tab:group_analysis}
\begin{tabular}{>{\centering\arraybackslash}C{1cm}c|cccc}
\specialrule{1pt}{2pt}{2pt}
\multicolumn{2}{c}{\textbf{Model}} &  \multicolumn{4}{c}{\textbf{LLaMA-3-8B}} \\
\specialrule{1pt}{2pt}{2pt}
\multicolumn{2}{c|}{\textbf{Configuration}} &  \multicolumn{4}{c}{\textbf{Group Size}} \\
\textbf{Method} & \textbf{SF/ZP} &\textbf{Per-Ch.} &\textbf{G128} & \textbf{G64} & \textbf{G32} \\
\specialrule{1pt}{2pt}{2pt}

Quarot & FP16  & 8.16 & 7.31 & 7.19 & 7.04 \\
DuQuant & FP16 & 8.06 & 8.02 & 8.03 & 7.98 \\
LightRot  & FP16    & -- & \textbf{6.99} & \textbf{6.87} & \textbf{6.80} \\
\rowcolor[HTML]{E0F2F5} 
\textbf{GyRot}  & FP16    & -- & \textbf{7.01} & \textbf{6.91} & \textbf{6.81} \\
\specialrule{0.5pt}{2pt}{2pt}

LightRot  & \textbf{INT8}    & -- & 7.69 & 36.33 & 30.12 \\
\rowcolor[HTML]{E0F2F5} 
\textbf{GyRot}  & \textbf{INT8}    & -- & \textbf{7.10} & \textbf{7.00} & \textbf{6.91} \\
\specialrule{1pt}{2pt}{0pt}

\end{tabular}
\vspace{-12pt}
\end{table}

\textcolor{black}{
\label{sec:low_bit_eval}
{\bf Evaluation in an extremely Low-bit Setting.}
To further validate the robustness of GyRot in the low-bit regime, we evaluate its performance under a more aggressive 3-bit weight quantization (W3A4 configuration). 
As shown in Table~\ref{tab:low_bit_analysis}, GyRot maintains strong performance even in this extremely low-bit condition, achieving comparable perplexity to LightRot and consistently outperforming Quarot across all model scales. 
In particular, GyRot-FP achieves 6.20/5.48 PPL on LLaMA-1-7B/13B, matching or surpassing prior rotation-based methods, while GyRot-INT shows only marginal degradation despite using fully integer scaling and zero-points. 
These results confirm that the cooperative rotation–group quantization design of GyRot remains effective even when bit precision is aggressively reduced, demonstrating its applicability beyond the standard 4-bit regime.
}

\begin{table}[t]
\centering
\caption{\textcolor{black}{Perplexity with Low-bit Quantization (W3A4).}}
\scriptsize
\label{tab:low_bit_analysis}
\begin{tabular}{>{\centering\arraybackslash}C{1.5cm}|ccccc}
\specialrule{1pt}{2pt}{2pt}
\multirow{2}{*}{\textbf{W3A4KV4}} & \multicolumn{5}{c}{\textbf{LLaMA}} \\
 & \textbf{1-7B} & \textbf{1-13B} & \textbf{2-7B} & \textbf{2-13B} & \textbf{3-8B} \\
\specialrule{1pt}{2pt}{2pt}

\textbf{Quarot} & 6.67  & 5.82 & 6.91 & 5.89 & 9.17 \\
\textbf{LightRot}  & 6.30    & 5.54 & \textbf{6.16} & \textbf{5.44} & 8.00 \\
\rowcolor[HTML]{E0F2F5} 
\textbf{GyRot-FP}  & \textbf{6.20}    & \textbf{5.48} & \textbf{6.16} & 5.48 & \textbf{7.73} \\
\rowcolor[HTML]{E0F2F5} 
\textbf{GyRot-INT}  & 6.22    & 5.49 & 6.64 & 5.50 & 7.83 \\
\specialrule{0.5pt}{2pt}{2pt}

\end{tabular}
\vspace{-8pt}
\end{table}

\subsection {Power, Performance and Area Evaluation}
\begin{figure}[]
\centering
\includegraphics[width=3.0in]{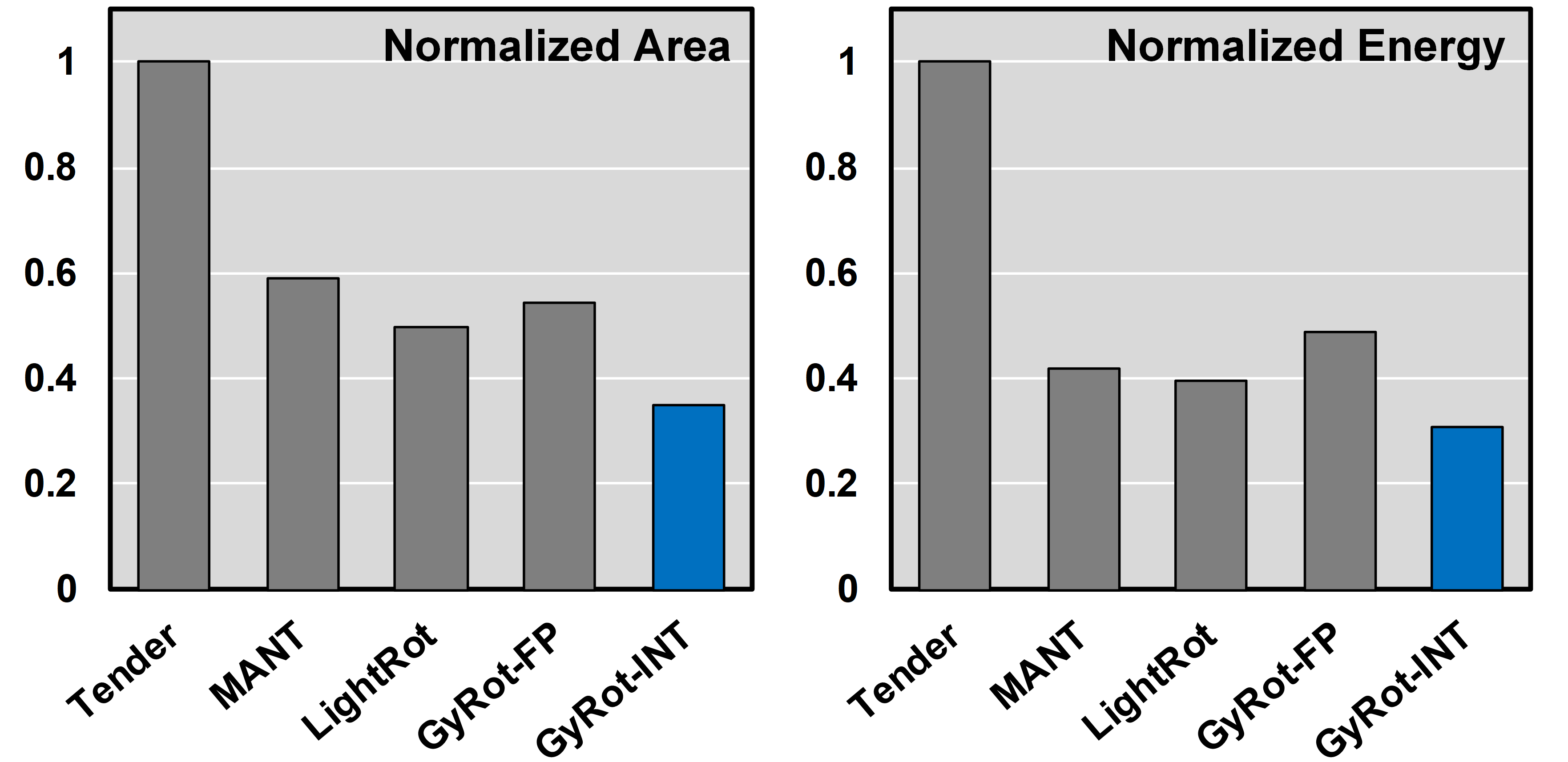}
\vspace{-6pt}
\caption{PE area and energy comparison}
\label{fig:f11}
\vspace{-12pt}
\end{figure}

\begin{figure*}[]
\centering
\includegraphics[width=6.8in]{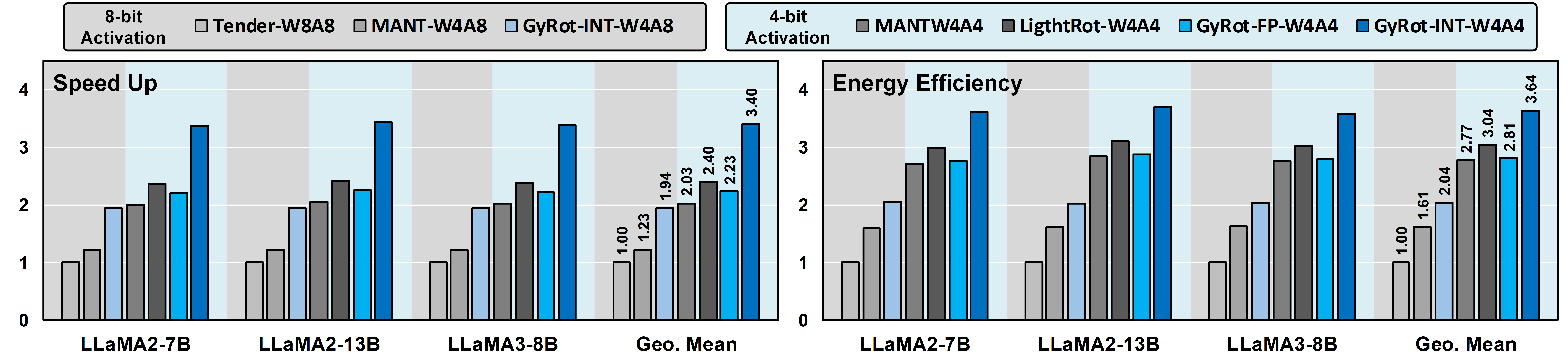}
\caption{Speedup and energy efficiency comparison across accelerators on WikiText2 with various bit configurations.}
\label{fig:f12}
\vspace{-12pt}
\end{figure*}

{\bf PE-level evaluation.}
Fig.~\ref{fig:f11} presents the normalized area and energy consumption of different LLM accelerators under iso-throughput conditions. All designs are synthesized in 28nm at 1GHz, based on the configuration detailed in Table~\ref{tab:perplexity_comparison}. For a fair comparison, Tender is modified to use dedicated 8-bit datapaths, removing any reconfiguration overhead associated with 4-bit operations. All other designs operate with 4-bit precision using group quantization, and their respective dequantization units are integrated into the PE based on group sizes: 128 for LightRot, 64 for MANT, and 32 for GyRot.
Tender, MANT, and LightRot adopt 2D systolic PE arrays, while \textit{GyRot} utilizes a 3D tensor PE array with systolic dataflow, as detailed in Section~V. Compared to Tender, \textit{GyRot-FP} achieves a 45.6\% area reduction and 51.0\% energy savings by leveraging low-bit group quantization combined with rotation for accuracy. However, the small group size of 32 increases dequantization overhead, resulting in higher cost compared to MANT and LightRot. MANT and LightRot employ group quantization with floating-point SF—LightRot additionally uses floating-point ZP—leading to extra hardware cost from FP arithmetic. By contrast, \textit{GyRot-INT} demonstrates the advantage of a fully integer implementation using INT8 SF and ZP, achieving the highest hardware efficiency with 65.2\% area and 69.2\% energy reduction over Tender.

{\bf System-level Performance and Energy Analysis.}
Fig.~\ref{fig:f12} presents the speedup and energy efficiency of {\it GyRot} compared to prior bit-flexible LLM accelerators under configurations that achieve similar accuracy levels, according to Table~\ref{tab:perplexity_comparison}. For 8-bit baselines, we compare against Tender-W8A8 and MANT-W4A8; for 4-bit settings, we include MANT-W4A4, LightRot-W4A4, and both FP and INT variants of GyRot-W4A4. 

Across all LLaMA models, {\it GyRot-INT} consistently outperforms existing methods in both performance and energy efficiency. It achieves a geometric mean speedup of $3.40\times$ and energy efficiency improvement of $3.64\times$ over the 8-bit Tender baseline. Compared to other group quantization baselines (MANT and LightRot), {\it GyRot-INT} delivers 41.7–67.5\% higher speedup and 19.8–31.4\% better energy efficiency on average. These gains stem from two sources: the use of a pure integer tensor PE without complex number formats like those in MANT, and a fully integer-based dequantization unit that avoids the overhead of floating-point scaling, biasing, and accumulation found in both LightRot and MANT.

\begin{figure}[]
\centering
\includegraphics[width=3.5in]{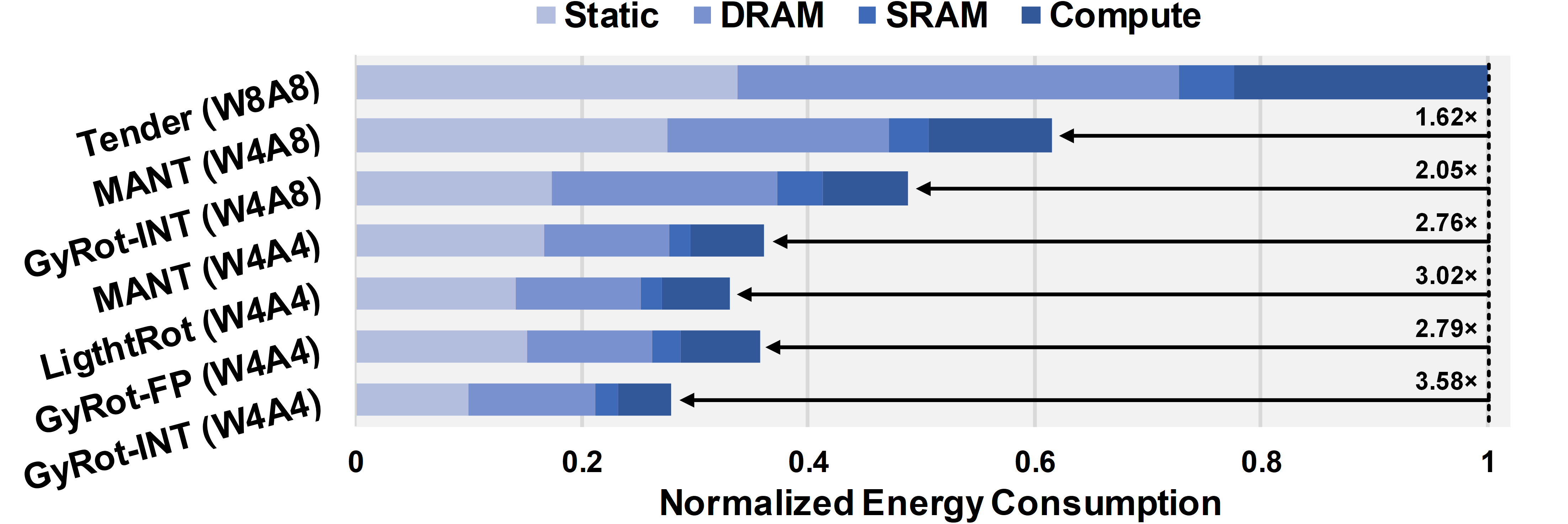}
\caption{Energy breakdown of GyRot in contrast with the baseline accelerators. Energy consumption with the LLaMA3-8B inference is evaluated.}
\label{fig:f13}
\vspace{-12pt}
\end{figure}

Fig.~\ref{fig:f13} shows the detailed energy breakdown when running LLaMA3-8B inference, categorized into static, DRAM, SRAM, and compute components. Compared to Tender, \textit{GyRot-INT} achieves substantial energy savings primarily through reduced DRAM access enabled by 4-bit operations and lower static power consumption resulting from its higher area efficiency and throughput. While prior 4-bit accelerators such as MANT and LightRot reduce compute energy via group quantization, they still incur significant energy and area overhead from floating-point dequantization. In contrast, \textit{GyRot-INT} leverages a fully integer-based dequantization pipeline to minimize this cost. Although the tensor PE architecture of GyRot slightly increases SRAM energy, this is outweighed by the greater reduction in compute energy, resulting in the lowest total energy consumption among all 4-bit accelerators.

\begin{table}[]
\centering
\scriptsize
\caption{Area and Power Breakdown of the Proposed Accelerator}
\label{tab:breakdown}
\begin{tabular}{llcc}
\specialrule{1pt}{2pt}{2pt}
\textbf{Component} & \textbf{Configuration} & \textbf{Area [mm\textsuperscript{2}]} & \textbf{Power [mW]} \\
\specialrule{1pt}{2pt}{2pt}
PE Array & 8×8×32 INT Tensor         & 0.26 (12.4\%)  & 410.24 (55.4\%) \\
PE Array & 8×8 Dequant. + Accum.     & 0.09 (4.2\%)   & 118.40 (16.0\%) \\
W\textunderscore{SUM} unit & 8×32-way Adder-Tree  & 0.01 (0.5\%)   & 17.09 (2.3\%) \\
Input Buf. & 64KB + 8KB (SF/ZP) & 0.24 (11.4\%)  & 82.43 (11.1\%) \\
Weight Buf.     & 64KB + 4KB (SF)     & 0.23 (10.8\%)  & 64.38 (8.7\%) \\
Global Buf.     & 512KB               & 1.20 (57.1\%)  & 41.63 (5.6\%) \\
OVU               & 32-way + FHT unit   & 0.07 (3.5\%)   & 6.78 (0.9\%) \\
\specialrule{0.5pt}{2pt}{2pt}
\textbf{Total} &                         & \textbf{2.10 (100.0\%)} & \textbf{740.95 (100.0\%)} \\
\specialrule{1pt}{2pt}{0pt}
\end{tabular}
\vspace{-12pt}
\end{table}
{\bf Power and Area Breakdown.}
Table~\ref{tab:breakdown} summarizes the area and power distribution of the proposed \textit{GyRot-INT} accelerator. The majority of the area is occupied by the 512KB global buffer (57.1\%), which is shared across the chip for activations and intermediate storage. The integer tensor PE accounts for 12.4\% of the area and more than half of the total power consumption (55.4\%), reflecting its central role in computation. The dequantization and accumulation logic, tightly integrated within each PE, introduces a modest overhead of 4.2\% in area and 16.0\% in power. The FVU, responsible for non-linear vector operations and rotation using FHT, contributes minimally to the overall cost, consuming only 3.5\% of the area and 0.9\% of the power.

\section{Conclusion}
In this work, we presented GyRot, a quantization framework and accelerator architecture that leverages the benefits of rotation-based and fine-grained group quantization for low-bit LLM inference. We identified the inherent conflict between global rotation and localized scaling, and proposed CoRFiG and HAP to align rotation with group structure while enhancing quantizability through harmonic-aware permutation. Additionally, we introduced a reformulated asymmetric quantization scheme and zero-point rounding policy that reduce hardware overhead and enable fully integer implementation. Experimental results demonstrate that GyRot achieves state-of-the-art 4-bit accuracy and consistent performance gains across perplexity, zero-shot tasks, and conversational benchmarks, while delivering up to 3.4× speedup and 3.6× energy efficiency over existing accelerators. These results validate GyRot as a practical and scalable solution for efficient LLM deployment.

\section*{Acknowledgments}
This work was supported by Institute of Information \& communications Technology Planning \& Evaluation (IITP) under the Graduate School of Artificial Intelligence Semiconductor(IITP-2025-RS-2023-00256472) grant funded by the Korea government(MSIT)

\bibliographystyle{IEEEtranS}
\bibliography{refs}

\end{document}